\documentclass[useAMS,usenatbib]{mn2e}
\usepackage{courier}
\usepackage{times}
\usepackage{chngpage}
\usepackage{color}
\usepackage{float}
\usepackage[hidelinks]{hyperref}
\usepackage[fleqn]{amsmath}
\usepackage{amssymb}
\usepackage{relsize}
 \usepackage{graphicx}
\usepackage{multirow}
\title[Main-belt comets in PTF]{Main-belt comets in the Palomar Transient Factory survey:\\ I. The search for extendedness}
\author[A. Waszczak et al.]{A. Waszczak,$^1$\thanks{E-mail: waszczak@caltech.edu} E. O. Ofek,$^2$ O. Aharonson,$^{1,3}$ S. R. Kulkarni,$^{1,4}$ D. Polishook,$^5$\newauthor J. M. Bauer,$^{6,7}$ D. Levitan,$^4$ B. Sesar,$^4$ R. Laher,$^8$ J. Surace,$^8$ and the PTF Team\\
\\$^1$Division of Geological and Planetary Sciences, California Institute of Technology, Pasadena, CA 91125, USA\\
         $^2$Benoziyo Center for Astrophysics, Faculty of Physics, Weizmann Institute of Science, Rehovot 76100, Israel\\
         $^3$Helen Kimmel Center for Planetary Science, Weizmann Institute of Science, Rehovot 76100, Israel\\
         $^4$Division of Physics, Mathematics and Astronomy, California Institute of Technology, Pasadena, CA 91125, USA\\
         $^5$Department of Earth, Atmospheric and Planetary Sciences, Massachusetts Institute of Technology, Cambridge, MA 02139, USA\\
         $^6$Jet Propulsion Laboratory, California Institute of Technology, Pasadena, CA 91109, USA\\
         $^7$Infrared Processing and Analysis Center, California Institute of Technology, Pasadena, CA 91125, USA\\
         $^8$Spitzer Science Center, California Institute of Technology, Pasadena, CA 91125, USA}
\begin{document}

\date{Accepted 2013 May 30.  Received 2013 May 26; in original form 2013 January 8}

\pagerange{\pageref{firstpage}--\pageref{lastpage}} \pubyear{2013}

\maketitle

\label{firstpage}

\begin{abstract}
Cometary activity in main-belt asteroids probes the ice content of these objects and provides clues to the history of volatiles in the inner solar system. We search the Palomar Transient Factory (PTF) survey to derive upper limits on the population size of active main-belt comets (MBCs). From data collected March 2009 through July 2012, we extracted $\sim$2 million observations of $\sim$220 thousand known main-belt objects (40\% of the known population, down to $\sim$1-km diameter) and discovered 626 new objects in multi-night linked detections. We formally quantify the ``extendedness'' of a small-body observation, account for systematic variation in this metric ({\it e.g.}, due to on-sky motion) and evaluate this method's robustness in identifying cometary activity using observations of 115 comets, including two known candidate MBCs and six newly-discovered non-main-belt comets (two of which were originally designated as asteroids by other surveys). We demonstrate a 66\% detection efficiency with respect to the extendedness distribution of the 115 sampled comets, and a 100\% detection efficiency with respect to extendedness levels greater than or equal to those we observed in the known candidate MBCs P/2010 R2 (La Sagra) and P/2006 VW$_{139}$. Using a log-constant prior, we infer 95\% confidence upper limits of 33 and 22 active MBCs (per million main-belt asteroids down to $\sim$1-km diameter), for detection efficiencies of 66\% and 100\%, respectively. In a follow-up to this morphological search, we will perform a photometric (disk-integrated brightening) search for MBCs.
\end{abstract}


\section{Introduction}

Though often regarded as quiescent rock- and dust-covered small bodies, asteroids can eject material by a variety of physical mechanisms. One subgroup of these \emph{active asteroids} \citep{jew12} are the \emph{main-belt comets} (MBCs), which we define\footnote{Some controversy surrounds the definitions of ``main-belt comet'', ``active asteroid'' and ``impacted asteroid''. While the term \emph{active main-belt object} (\citealp{bau12}; \citealp{ste12}) is the most general, our particular definition and usage of \emph{main-belt comet} is intended to follow that of \cite{hsi06a}, {\it i.e.}, periodic activity due to sublimating volatiles.} as objects in the dynamically-stable main asteroid belt that exhibit a periodic ({\it e.g.}, near-perihelion) cometary appearance due to the sublimation of freshly collisionally-excavated ice. Prior to collisional excavation, this ice could persist over the age of the solar system, even in the relatively warm vicinity of $\sim$3 AU, if buried under a sufficiently thick layer of dry porous regolith (\citealp{sch08}; \citealp{pri09}).
\begin{table*}
\caption{Known candidate MBCs and impacted asteroids (collectively termed ``active main-belt objects'') as of April 2013: Summary of orbits and sizes}
\hfill{}
\begin{tabular}{llccccll}
\hline
main-belt comets&asteroid&$a$ (AU)&$e$&$i$ (deg)&$D$ (km)&perihelia&references\\
\hline
133P/Elst-Pizarro&7968&3.157&0.16&1.4&$3.8\pm0.6$&'96,'02,'07&Elst+ '96\nocite{els96}; Marsden '96; Hsieh+ '04,'09a,'10\\
176P/LINEAR&118401&3.196&0.19&0.2&$4.0\pm0.4$&'05,'11&Hsieh+ '06b,'09a,'11a; Green '06\\
238P/Read& &3.165&0.25&1.3&$\sim$0.8&'05,'10&Read '05; Hsieh+ '09b,'11b\\
259P/Garradd& &2.726&0.34&15.9&$0.3\pm0.02$&'08&Garradd+ '08\nocite{gar08}; Jewitt+ '09; MacLennan+ '12\\
P/2010 R2 (La Sagra)& &3.099&0.15&21.4&$\sim$1.4&'10&Nomen+ '10; Moreno+ '11; Hsieh+ '12a\\
P/2006 VW$_{139}$&300163&3.052&0.20&3.2&$\sim$3.0&'11&Hsieh+ '12b; Novakovi\'c+ '12; Jewitt+ '12\\
P/2012 T1 (PANSTARRS)& &3.047&0.21&11.4&$\sim$2&'12&Hsieh+ '12c\\
\hline
impacted asteroids& & & & & & & \\
\hline
P/2010 A2 (LINEAR)& &2.291&0.12&5.3&$\sim$0.12& &Birtwhistle+ '10; Jewitt+ '10; Snodgrass+ '10\\
Scheila&596&2.927&0.17&14.7&$113\pm 2$& &Larson '10; Jewitt+ '11; Bodewitts+ '11\\
P/2012 F5 Gibbs& &3.004&0.04&9.7&$<2.1$& &Gibbs+ '12; Stevenson+ '12; Moreno+ '12\\
\hline
\end{tabular}
\hfill{}
\end{table*}
\\ \indent More complete knowledge of the number distribution of ice-rich asteroids as a function of orbital ({\it e.g.}, semi-major axis) and physical ({\it e.g.}, diameter) properties could help constrain dynamical models of the early solar system (\citealp{morb12} and refs. therein).  Such models trace the evolution of primordially distributed volatiles, including the ``snow line'' of H$_2$O and other similarly stratified compounds. Complemented by cosmochemical and geochemical evidence ({\it e.g.}, \citealp{owe08}; \citealp{alb09}; \citealp{rob11}), such models explore the possibility of late-stage (post-lunar formation) accretion of Earth's and/or Mars' water from main-belt objects. Some dynamical simulations (\citealp{lev09}; \citealp{wal11}) suggest that emplacement of outer solar system bodies into the main asteroid belt may have occurred; these hypotheses can also be tested for consistency with a better-characterized MBC population. 

For at least the past two decades ({\it e.g.}, \citealp{luu92}), visible band CCD photometry has been regarded as a viable means of searching for subtle cometary activity in asteroids---spectroscopy being an often-proposed alternative.  However, existing visible spectra of MBCs are essentially indistinguishable from those of neighboring asteroids. Even with lengthy integration times, active MBC spectra in the UV and visible lack the bright 388-nm cyanogen (CN) emission line seen in conventional comets \citep{lic11}. Near-infrared MBC spectra are compatible with water ice-bearing mixtures of carbon, silicates and tholins but also suffer very low signal-to-noise \citep{rou11}. The larger asteroids Themis (the likely parent body of several MBCs) and Cybele show a 3-$\rm{\mu}$m absorption feature compatible with frost-covered grains, but the mineral goethite could also produce this feature (\citealp{jewg12} and refs. therein). The \emph{Herschel Space Observatory} targeted one MBC in search of far-infrared H$_2$O-line emission, yet only derived an upper limit for gas production \citep{dev12}. In general, the low albedo \citep{bau12} and small diameter of MBCs ($\sim$km-scale), along with their low activity relative to conventional comets, makes them unfit for spectroscopic discovery and follow-up. Imaging of their sunlight-reflecting dust and time-monitoring of disk-integrated flux, however, are formidable alternatives which motivate the present study.

There are seven currently known candidate MBCs (Table 1) out of $\sim$560,000 known main-belt asteroids. These seven are regarded as candidates rather than true MBCs because they all lack direct evidence of constituent volatile species, although two (133P and 238P) have shown recurrent activity at successive perihelia. Three other active main-belt objects---P/2010 A2 (LINEAR), 596 Scheila, and P/2012 F5 (Gibbs)---likely resulted from dry collisional events and are thus not considered to be candidate MBCs. Four of the seven MBCs were discovered serendipitously by individuals or untargeted surveys. The other three were found systematically: the first in the \emph{Hawaii Trails Project} \citep{hsi09}, in which targeted observations of $\sim$600 asteroids were visually inspected , and the latter two during the \emph{Pan-STARRS 1} (PS1; \citealp{kai02}) survey, by an automated point-spread function analysis subroutine in the PS1 moving-object pipeline \citep{hsi12b}. Three of the seven candidate MBCs were originally designated as asteroids, including two of the three systematically discovered ones, which were labeled as asteroids for more than five years following their respective discoveries by the automated NEO surveys LINEAR \citep{sto00} and Spacewatch (\citealp{geh84}; \citealp{mcm00}).

Prior to this work, two additional untargeted MBC searches have been published. Gilbert and Wiegert (\citeyear{gil09}, \citeyear{gil10}) checked 25,240 moving objects occurring in the \emph{Canada-France-Hawaii Telescope Legacy Survey} \citep{jon06} using automated PSF comparison against nearby field stars and visual inspection. Their sample, consisting of both known and newly-discovered objects extending down to a limiting diameter of $\sim$1-km, revealed cometary activity on one new object, whose orbit is likely that of a Jupiter-family comet. \cite{son11} analyzed 924 asteroids (a mix of known and new, down to $\sim$0.5-km diameter) observed in the \emph{Thousand Asteroid Light Curve Survey} \citep{mas09}. They fit stacked observations to model comae and employed a tail-detection algorithm. While their sample did not reveal any new MBCs, they introduced a solid statistical framework for interpreting MBC searches of this kind, including the proper Bayesian treatment of a null-result.

In this work, we first describe the process of extracting observations of known and new solar system small bodies in the PTF survey. We next establish a metric for ``extendedness'' and a means of correcting for systematic (non-cometary) variation in this metric. We then apply this metric to a screening process wherein individual observations are inspected by eye for cometary appearance. Finally, we apply our results to upper limit estimates of the population size of active main-belt comets.

\section{Raw transient data}

\subsection{Survey overview}

The \emph{Palomar Transient Factory} (PTF)\footnote{\href{http://ptf.caltech.edu}{\color{blue}{http://ptf.caltech.edu}}} is a synoptic survey designed to explore the transient and variable sky (\citealp{law09}; \citealp{rau09}). The PTF camera, mounted on Palomar Observatory's 1.2-m $f/2.44$ Oschin Schmidt Telescope, uses 11 CCDs ($4096\times2048$ each) to observe 7.26 deg$^2$ of the sky at a time with a resolution of $1.01''$/pixel. Most exposures use either a Mould-$R$ or Gunn-$g'$ filter and are 60-s (a small fraction of exposures also comprise an H$\alpha$-band survey of the sky). Science operations began in March 2009, with a nominal 2- to 5-day cadence for supernova discovery and typically twice-per-night imaging of fields. Median seeing is $2''$ with a limiting apparent magnitude $R\sim20.5$ (5$\sigma$), while near-zenith pointings under dark conditions routinely achieve $R\sim21.0$ \citep{law10}. 

PTF pointings (Figure 1) and cadences are not deliberately selected for solar system science. In fact, PTF's routine sampling of high ecliptic latitudes (to avoid the sometimes bright Moon) alleviates small-body sampling bias with respect to orbital inclination (see Section 3.3).

\begin{figure}
\centering
\includegraphics[scale=0.16]{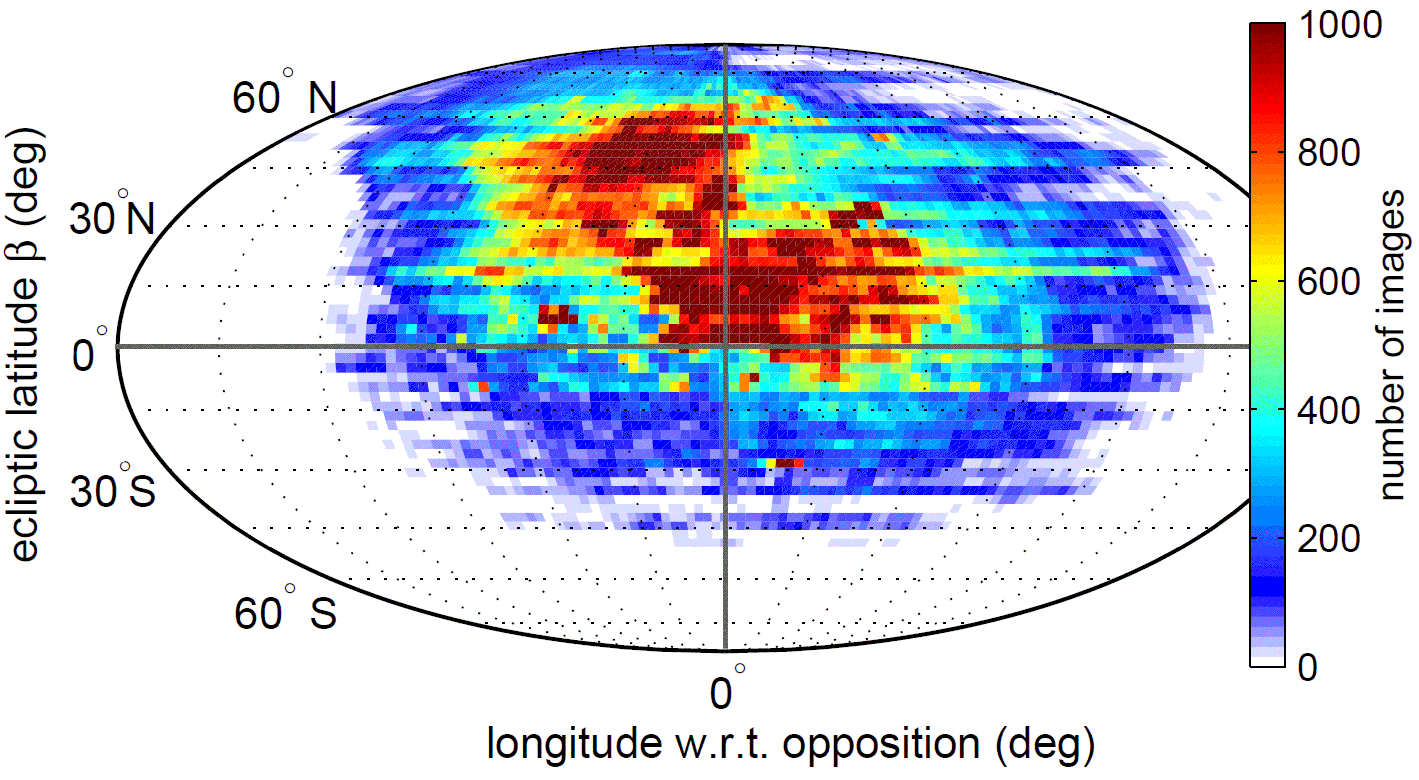}
\caption{Distribution of PTF pointings over the first 41 months of operations (March 2009 through July 2012), in sky coordinates relevant to small-body observations.}
\end{figure}

We use data that have been reduced by the PTF photometric pipeline (\citealp{gri10}; Laher al. in prep) hosted at the \emph{Infrared Processing and Analysis Center} (IPAC) at Caltech. For each image, the pipeline performs debiasing, flat-fielding, astrometric calibration, generation of mask images, and creation of a catalog of point sources using the astrometric reduction software SExtractor \citep{ber96}. Code-face parameters such as \texttt{MAGERR\_AUTO} in this work refer to SExtractor output quantities.

Absolute photometric calibration is described in Ofek et al. (\citeyear{ofe12a}, \citeyear{ofe12b}) and routinely achieves precision of $\sim$0.02 mag under photometric conditions. In this work, we use relative (lightcurve-calibrated) photometry (Levitan et al. in prep; for algorithm details see \citealp{lev11}), which has systematic errors of 6--8 mmag in the bright (non-Poisson-noise-dominated) regime. Image-level (header) data used in this study were archived and retrieved using an implementation of the Large Survey Database software (LSD, \citealp{jur11}), whereas detection-level data were retrieved from the PTF photometric database.

\subsection{Candidate-observation quality filtering}

Prior to ingestion into the photometric database, individual sources are matched against a PTF reference image (a deep co-add consisting of at least $\sim$20 exposures, reaching $\ge21.7$ mag). Any detection not within $1.5''$ of a reference object is classified in the database as a transient. The ensemble of transients forms a raw sample from which we seek to extract asteroid (and potential MBC) observations. As of 2012-Jul-31 there exist $\sim$30 thousand deep reference images (unique filter-field-chip combinations) against which $\sim$1.6 million individual epoch images have been matched, producing a total of $\sim$700 million transients. Of these, we discard transients which satisfy any of the following constraints:

\begin{itemize} \itemsep2pt
\item within $4''$ of a reference object
\item outside the convex footprint of the reference image
\item from an image with astrometric fit error $>1''$ relative to the 2MASS survey \citep{skr06} or systematic relative photometric error $>$ 0.1 mag
\item within 6.5 arcmin of a $V<7$ Tycho-2 star \citep{hog00}, the approximate halo radius of very bright stars in PTF
\item within 2 arcmin of either a $7<V<10$ Tycho-2 star or a $7<R<10$ PTF reference source, a lower-order halo radius seen in fainter stars
\item within 1 arcmin of a $10<R<13$ PTF reference source; most stars in this magnitude range do not have halos but do have saturation and blooming artifacts
\item within 30 pixel-columns of a $V<10$ Tycho-2 star on the same image (targets blooming columns)
\item within 30 pixels of the CCD edge
\item flagged by the IPAC pipeline as either an aircraft/satellite track, high dark current pixel, noisy/hot pixel, saturated pixel, dead/bad pixel, ghost image, dirt on the optics, CCD-bleed or bright star halo (although the above-described bright star masks are more aggressive than these last two flags)
\item flagged by SExtractor as being either photometrically unreliable due to a nearby source, originally blended with another source, saturated, truncated or processed during a memory overflow
\item overconcentrated in flux relative to normal-PSF (stellar) objects on the image ({\it i.e.} single-pixel radiation hit candidates)---true if the source's \texttt{MU\_MAX $-$ MAG\_AUTO} value minus the image's median stellar \texttt{MU\_MAX $-$ MAG\_AUTO} value is less than $-1$ (this criterion is further explained in Section 5)
\end{itemize}

Application of the above filtering criteria reduces the number of transients (moving-object candidates) from $\sim$700 million to $\sim$60 million detections. While greatly reduced, this sample size is still too large to search (via the methods outlined in the following sections) given available computing resources---hence we seek to further refine it.  These non-small-body detections are likely to include random noise, difficult-to-flag ghost features \citep{yan02}, less-concentrated radiation hits, bright star and galaxy features missed by the masking process, clouds from non-photometric nights, and real astrophysical transients ({\it e.g.}, supernova). 

We find that about two-thirds of the transients in this sample occur in the densest $\sim$10\% of the images ({\it i.e.} images with more than $\sim$50 transients). These densest $\sim$10\% of images represent over 50\% of all images on the galactic equator ($|b|<20^{\circ}$), but only 7\% of all images on the ecliptic ($|\beta|<20^{\circ}$). Hence, discarding them from our sample should not have a significant effect on the number of small-body observations we extract. Discarding these dense images reduces our sample of transients to $\sim$20 million.

\subsection{Sample quality assessment}

\begin{figure}
\centering
\includegraphics[scale=0.13]{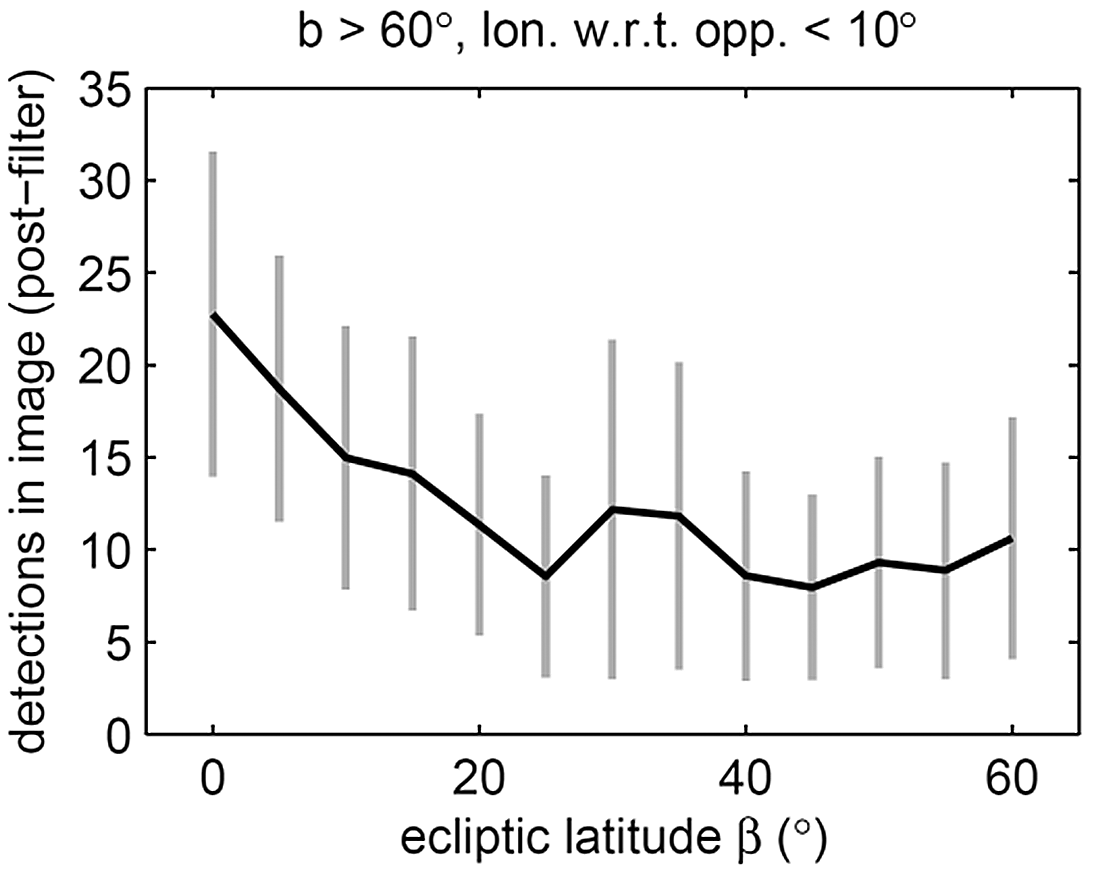}
\caption{Transients per image (after filtering and discarding the densest 10\% of images) versus ecliptic latitude, off the galactic equator and near-opposition longitude. Vertical gray lines are the scatter (standard deviation) and the black line traces the mean number of detections. Due to the large number of images, the standard error of the mean for each bin is very small (comparable to the width of the black line). The inferred ratio of false positive detections (image artifacts) to real small-body detections at low ecliptic latitudes is at least of order unity.}
\end{figure}

Figure 2 details the distribution of transients (candidates) per image as a function of ecliptic latitude, after applying the above filters and discarding the dense images. The galactic signal (not shown) is still present: off-ecliptic low galactic latitude fields have a mean of $\sim$40 transients; this number drops roughly linearly with galactic latitude, implying significant residual contribution from ghosts and other missed dense-field artifacts. However, off the galactic equator a factor-of-two increase in the mean number of detections per image is seen from $|\beta|=50^{\circ}$ toward the ecliptic, indicating a clear detection of the solar system's main belt. 

\section{Known-object extraction}

Having defined our sample of candidate observations, we now seek to match it to objects with known orbits. We first index the candidate observations into a three-dimensional kd-tree, then match this tree against ephemeris data (predicted positions) for all objects. The reader who wishes to skip over the details of the matching algorithm should now go to Section 3.3.

\subsection{Implementation of kd-tree indexing}

A kd-tree (short for $k$-dimensional tree) is a data structure which facilitates efficient cross-matching of $M$ query points against $N$ data points via a multi-dimensional binary search. Whereas a brute force cross-matching involves of order $MN$ computations, a kd-tree reduces this to order $M\log{⁡N}$. \cite{kub07} gives an introduction to kd-trees (including some terminology we use below) and details their increasingly common application in the moving-object processing subsystem (MOPS) of modern sky surveys.

Our kd-tree has the following features. Since the detections are three dimensional points (two sky coordinates plus one time), the tree's \emph{nodes} are box volumes, each of which is stored in memory as six double precision numbers. Before any leaf nodes (single datum nodes) are reached, the $n^{\rm{th}}$  level of the tree consists of 2$^{n-1}$ nodes, hence each level of the tree is stored in an array of size 2$^{n-1}\times 6$ or smaller.

After definition via median splitting, the bounds of each node are set to those of the smallest volume enclosing all of its data. The splitting of nodes is a parallelized component of the tree-construction algorithm (which is crucial given their exponential increase in number at each successive level). Because the splitting-dimension is cycled continuously, the algorithm will eventually attempt to split data from a single image along the time dimension; when this occurs it simply postpones splitting until the next level (where it is split spatially).

\subsection{Matching ephemerides against the kd-tree}

After constructing the kd-tree of moving-object candidates, we search the tree for known objects. For each of the $\sim$600 thousand known solar system small bodies we query JPL's online ephemeris generator HORIZONS \citep{gio96} to produce a one-day spaced ephemeris over the 41-month time span of our detections (2009-Mar-01 to 2012-Jul-31). We then search this ephemeris against the kd-tree of candidate detections. In particular, the 1,250 points (days) comprising the ephemeris are themselves organized into a separate (and much smaller) kd-tree-like structure, whose nodes are instead defined by splits exclusively in the time dimension and whose leaf nodes always consist of two ephemeris points spaced one day apart.

The ephemeris tree is ``pruned'' as it is grown, meaning that at each successive level all ephemeris nodes not intersecting at least one PTF tree node (at the same tree level) are discarded from the tree. Crossing of the $\rm{R.A.}=0^{\circ}$ discontinuity is dealt with by detecting nodes that span nearly $360^{\circ}$ in R.A. at sufficiently high tree levels. To account for positional uncertainty in the ephemeris, each ephemeris node is given an $8''$ buffer in the spatial dimensions, increasing its volume slightly and ensuring that the ephemeris points themselves never lie exactly on any of the node vertices.

Once the ephemeris tree is grown to only leaf nodes (which are necessarily overlapping some PTF transients), HORIZONS is re-queried for the small body's position at all unique transient epochs found in each remaining one-day node. Since each leaf node's angular footprint on the sky is of order the square of the object's daily motion ($\sim$10 arcmin$^2$ for main-belt objects---much less than the size of a PTF image), the number of unique epochs is usually small, on the order of a few to tens. The PTF-epoch-specific ephemerides are then compared directly with the handful of candidate detections in the node, and matches within $4''$ are saved as confirmed small-body detections. In addition to the astrometric and photometric data from the PTF pipeline, orbital geometry data from HORIZONS are saved. 

\begin{figure*}
\centering
\includegraphics[scale=0.26]{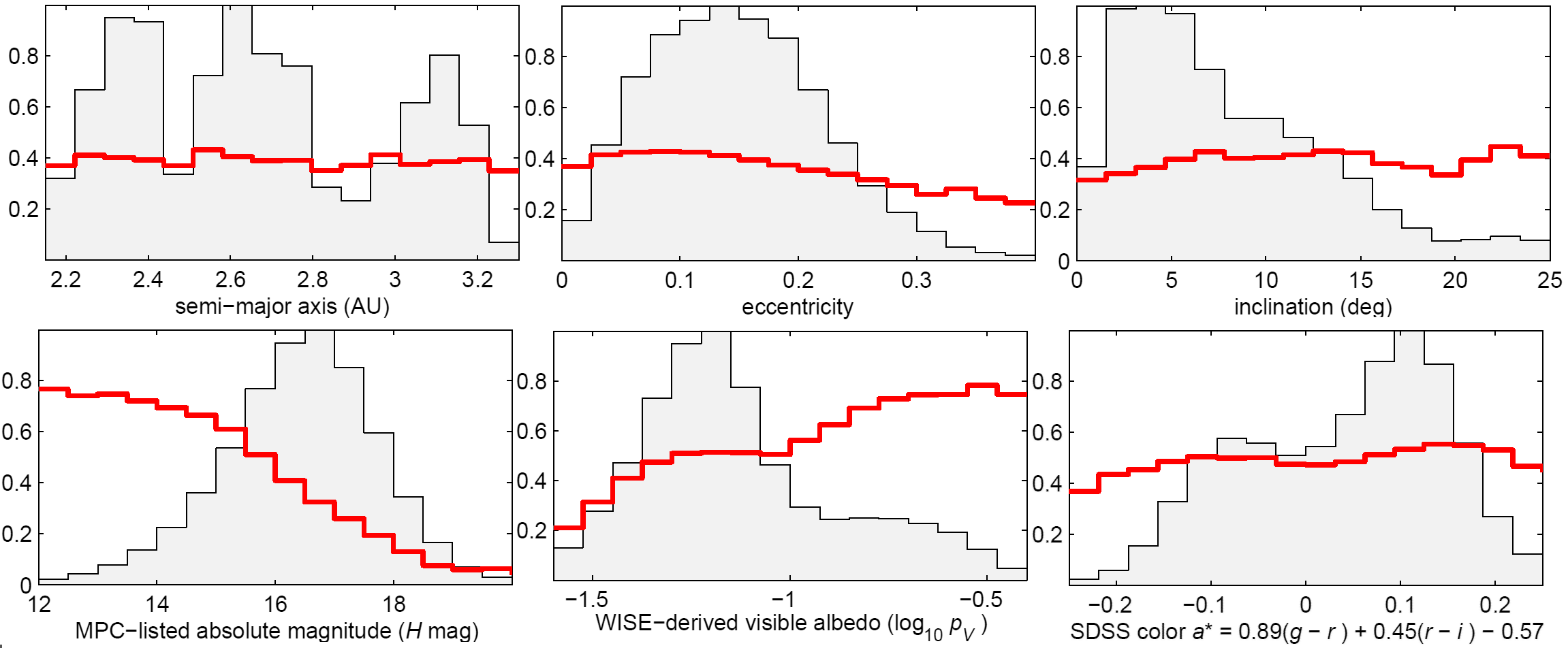}
\caption{The shaded gray histograms show the distributions of known objects (normalized such that largest bin equals unity), while the red lines show the fraction of objects in each bin included in the PTF dataset. The osculating orbital elements are from the JPL Small-Body Database, (\href{http://ssd.jpl.nasa.gov}{\color{blue}{http://ssd.jpl.nasa.gov}}), the absolute magnitudes from the Minor Planet Center (\href{http://www.minorplanetcenter.net}{\color{blue}{http://www.minorplanetcenter.net}}), the visible albedos from fits to the WISE cryogenic data \citep{mas11}, and the Sloan colors from the SDSSMOC 4th release \citep{par08} supplemented with 2008--2009 data (B. Sesar, personal communication).}
\end{figure*}

Given the candidate sample of $\sim$20 million transients, for each known object the search takes $\sim$4 seconds (including the HORIZONS queries, the kd-tree search and saving of confirmed detections). Hence, PTF observations of the $\sim$600 thousand known small bodies (main-belt objects, near-Earth objects, trans-Neptunian objects, comets, etc.) require $\sim$4 days to harvest on an 8-core machine. This relatively quick run time is crucial given that both the list of PTF transients and the list of known small bodies are updated regularly, necessitating periodic re-harvesting. 

\subsection{Summary of known small-bodies detected}

We used the known small bodies list current as of 2012-Aug-10, consisting of 333,841 numbered objects, 245,696 unnumbered objects, and 3157 comets (including lettered fragments and counting only the most recent-epoch orbital solution for each comet). Our search found 2,013,279 observations of 221,402 known main-belt objects in PTF ($\sim$40\% of all known). Table 2 details the coverage into various other orbital subpopulations.

Two active known candidate main-belt comets appeared in the sample: P/2010 R2 (La Sagra) was detected 34 times in 21 nights between 2010-Jul-06 and 2010-Oct-29, and P/2006 VW$_{139}$ was detected 5 times in 3 nights---2011-Sep-27, 2011-Oct-02 and 2011-Dec-21 (see Figure 13 in Section 6). In addition to these MBCs, there were 108 known Jupiter-family comets and 65 long-period comets in this sample (see Table 3).

\begin{figure}
\centering
\includegraphics[scale=0.12]{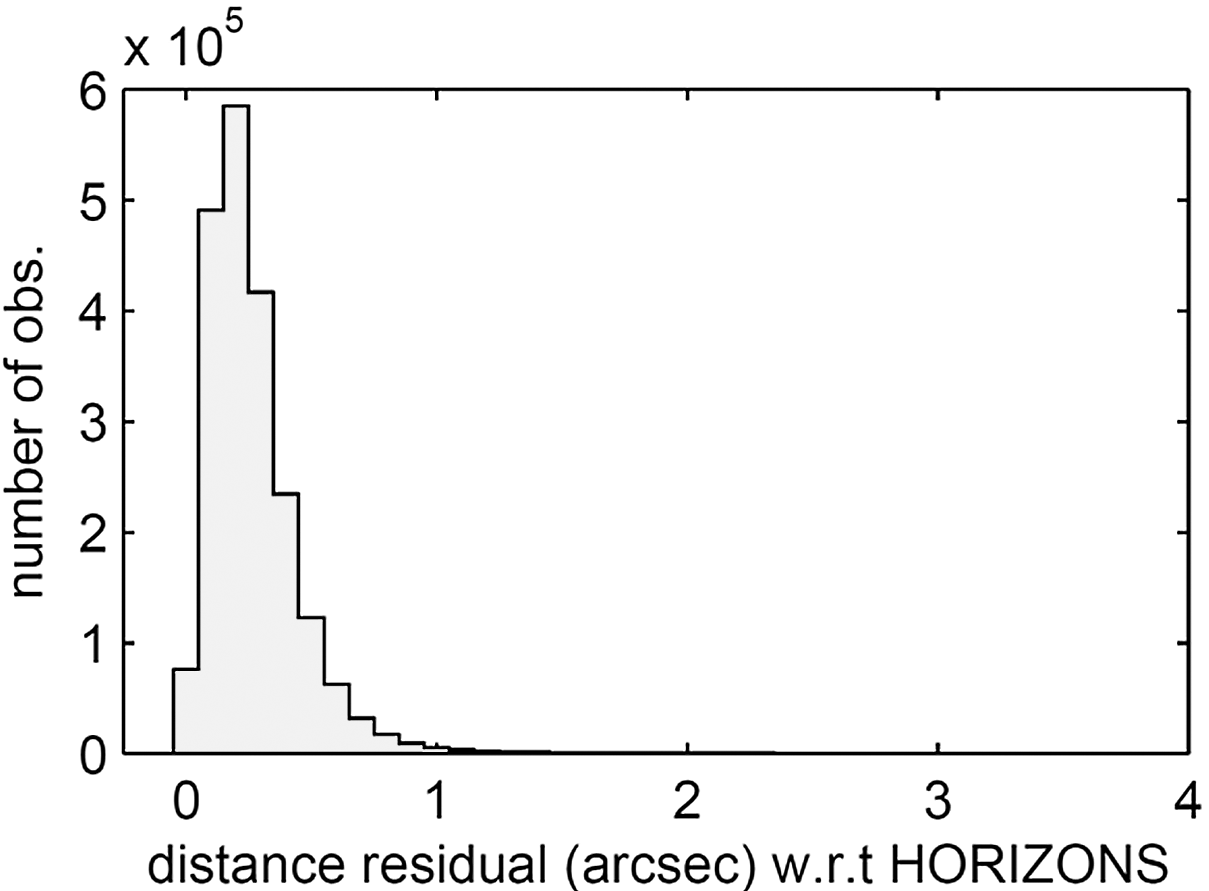}
\caption{Distance residuals of harvested small-body observations with respect to their predicted position. The horizontal axis intentionally extends to $4''$, as this is the matching radius we use. Significant contamination due to false-positive detections would increase with distance; this does not appear to be the case.}
\end{figure}

In terms of coverage, 54\% of the objects are observed five times or fewer, and 53\% of the objects are observed on three nights or fewer. Observation-specific statistics on this data set, such as apparent magnitudes, on-sky motions, etc., appear later in Section 5 (see the histograms in Figure 9). Lastly, a summary of orbital-coverage statistics appears in Figure 15 (Section 7).

The orbital distribution of the PTF sample is shown in the top row of Figure 3. The fraction of known objects sampled appears very nearly constant at 40\% across the full main-belt ranges of the orbital elements $a$, $e$ and $i$. With respect to absolute magnitude (referenced for all objects from the Minor Planet Center), the PTF sampling fraction of 40\% applies to the $H\sim17$ mag bin, corresponding to ~1-km diameter objects for a typical albedo of $\sim$10\%. 

As shown in Figure 4, the distribution of astrometric residuals with respect to the ephemeris prediction is sharply concentrated well within the matching threshold of $4''$. Were this data set significantly contaminated by randomly distributed false-positives, then their number would increase with matching distance ({\it i.e.}, with annular area per unit matching radius), which evidently is not the case.

\subsection{Overlap of PTF with the WISE and SDSS data sets}

During its full-cryogenic mission in 2010, the \emph{Wide-field Infrared Survey Explorer} (WISE; \citealp{wri10}; Masiero et al. \citeyear{mas11}, \citeyear{mas12}; \citealp{mai12}), observed 94,653 asteroids whose model-derived visible albedos ($p_V$) have errors of less than 0.05, and nearly half (45,321) of these were also observed by PTF. Relative to this known-albedo sample, PTF detected 47\% of the dark ($p_V<0.1$) and $\sim$69\% of the bright ($p_V>0.1$) asteroids (Figure 3, middle bottom). 

\begin{table}
\renewcommand{\arraystretch}{1.2}
\caption{Known solar system small-body detections in PTF}
\begin{tabular}{lccccc}
\hline
&\multirow{2}{*}{main-belt}&Trojan&\multirow{2}{*}{comets$^*$}&\multirow{2}{*}{NEOs}&TNOs \& \\[-0.8ex]
&                                           &\& Hilda&                                        &                                      &centaurs\\
\hline
detections&2,013,279&50,056&2,181&6,586&790\\
objects&221,402&5,259&175$^{\dagger}$&1,257&75\\
\% of known&39\%&55\%&3\%&13\%&4\%\\
\hline
\end{tabular}
\smallskip
\\$^*$See Table 5 for a more detailed breakdown by comet dynamical type.
\\$^{\dagger}$The count of 175 comets given here differs from the count of 115 given in the abstract, for various reasons described in Section 5.4 and Table 5.
\hfill{}
\end{table}

\begin{table*}
\caption{Known comets observed by PTF. Number of observations and nights; magnitude ranges, heliocentric ($r$) and geocentric ($\Delta$) distances (in AU).}
\hfill{}
\begin{tabular}{lrrccrrrrrr}
\hline\\[-3ex]
name&obs.&nights&first date&last date& $V_\text{min}$& $V_\text{max}$& $r_\text{min}$& $r_\text{max}$& $\Delta_\text{min}$& $\Delta_\text{max}$\\
\hline
7P/Pons-Winnecke                 &   5 &   3 & 2009-09-15 & 2009-09-21 & 20.3 & 21.0 &  3.5 &  3.5 &  2.9 &  2.9\\
9P/Tempel 1                      &   1 &   1 & 2010-03-16 & 2010-03-16 & 20.0 & 20.0 &  2.9 &  2.9 &  2.3 &  2.3\\
19P/Borrelly                     &   3 &   3 & 2009-05-14 & 2009-06-25 & 15.6 & 17.4 &  3.1 &  3.4 &  2.6 &  3.3\\
29P/Schwassmann-Wachmann 1       &  15 &  13 & 2011-01-10 & 2011-02-14 & 14.4 & 15.5 &  6.2 &  6.2 &  5.3 &  5.7\\
30P/Reinmuth 1                   &   1 &   1 & 2010-02-26 & 2010-02-26 & 15.3 & 15.3 &  1.9 &  1.9 &  1.5 &  1.5\\
31P/Schwassmann-Wachmann 2       &  26 &  15 & 2009-12-17 & 2011-04-11 & 18.1 & 19.0 &  3.5 &  3.6 &  2.5 &  2.8\\
33P/Daniel                       &   1 &   1 & 2009-03-18 & 2009-03-18 & 16.8 & 16.8 &  2.8 &  2.8 &  2.0 &  2.0\\
36P/Whipple                      &  45 &  14 & 2011-09-05 & 2011-11-22 & 17.9 & 19.4 &  3.1 &  3.1 &  2.1 &  2.5\\
47P/Ashbrook-Jackson             &   4 &   3 & 2009-12-03 & 2009-12-11 & 17.3 & 17.6 &  3.3 &  3.3 &  2.3 &  2.4\\
48P/Johnson                      &   5 &   3 & 2010-04-11 & 2011-06-11 & 16.2 & 21.4 &  2.4 &  3.8 &  1.6 &  2.8\\
49P/Arend-Rigaux                 &  30 &  12 & 2012-01-20 & 2012-07-16 & 16.0 & 20.0 &  1.7 &  2.9 &  1.0 &  3.1\\
54P/de Vico-Swift-NEAT           &   3 &   2 & 2009-07-23 & 2009-07-29 & 20.1 & 20.7 &  2.4 &  2.4 &  1.4 &  1.4\\
59P/Kearns-Kwee                  &   1 &   1 & 2010-02-17 & 2010-02-17 & 20.5 & 20.5 &  3.4 &  3.4 &  2.5 &  2.5\\
64P/Swift-Gehrels                &   3 &   3 & 2009-10-03 & 2010-02-14 & 15.8 & 18.6 &  1.9 &  2.9 &  1.9 &  2.2\\
65P/Gunn                         &   9 &   4 & 2009-05-25 & 2012-02-05 & 14.0 & 19.7 &  2.9 &  4.1 &  2.5 &  4.4\\
71P/Clark                        &   8 &   5 & 2011-01-12 & 2011-01-27 & 19.6 & 20.6 &  3.0 &  3.1 &  2.2 &  2.4\\
74P/Smirnova-Chernykh            &   6 &   3 & 2010-02-16 & 2010-02-23 & 16.1 & 16.5 &  3.6 &  3.6 &  3.0 &  3.1\\
77P/Longmore                     &  74 &   4 & 2009-03-13 & 2009-04-06 & 14.2 & 14.7 &  2.4 &  2.4 &  1.4 &  1.5\\
78P/Gehrels 2                    &  11 &   3 & 2011-11-02 & 2011-11-09 & 12.3 & 12.5 &  2.1 &  2.1 &  1.3 &  1.3\\
94P/Russell 4                    &   5 &   4 & 2010-02-16 & 2010-06-03 & 16.2 & 17.5 &  2.3 &  2.3 &  1.3 &  2.1\\
103P/Hartley 2                   &  18 &  11 & 2010-06-06 & 2010-08-03 & 14.6 & 18.3 &  1.5 &  2.1 &  0.7 &  1.5\\
10P/Tempel 2                     &   6 &   5 & 2009-06-09 & 2010-08-10 & 10.6 & 20.0 &  1.5 &  3.4 &  0.7 &  3.3\\
116P/Wild 4                      &   5 &   3 & 2009-03-27 & 2009-04-01 & 13.6 & 14.0 &  2.3 &  2.3 &  1.6 &  1.6\\
117P/Helin-Roman-Alu 1           &  29 &  13 & 2010-02-13 & 2011-12-04 & 18.8 & 19.8 &  4.5 &  5.1 &  4.5 &  4.8\\
118P/Shoemaker-Levy 4            &   3 &   2 & 2010-02-18 & 2010-04-08 & 13.6 & 14.7 &  2.0 &  2.1 &  1.3 &  1.9\\
123P/West-Hartley                &   1 &   1 & 2010-09-13 & 2010-09-13 & 20.3 & 20.3 &  3.0 &  3.0 &  2.9 &  2.9\\
127P/Holt-Olmstead               &   5 &   3 & 2009-08-13 & 2009-11-17 & 17.1 & 18.7 &  2.2 &  2.2 &  1.3 &  1.6\\
130P/McNaught-Hughes             &   2 &   2 & 2010-04-11 & 2010-04-16 & 20.8 & 21.0 &  3.5 &  3.5 &  2.5 &  2.5\\
131P/Mueller 2                   &   8 &   2 & 2011-11-02 & 2011-11-03 & 18.6 & 19.0 &  2.5 &  2.5 &  1.6 &  1.6\\
142P/Ge-Wang                     &   3 &   2 & 2010-10-03 & 2010-10-17 & 20.6 & 20.6 &  2.7 &  2.7 &  1.7 &  1.8\\
143P/Kowal-Mrkos                 &   2 &   1 & 2010-08-24 & 2010-08-24 & 19.8 & 20.1 &  3.7 &  3.7 &  2.9 &  2.9\\
149P/Mueller 4                   &  17 &  10 & 2010-02-16 & 2010-06-03 & 18.7 & 20.1 &  2.7 &  2.7 &  1.8 &  2.2\\
14P/Wolf                         &   1 &   1 & 2009-11-03 & 2009-11-03 & 19.4 & 19.4 &  3.1 &  3.1 &  2.2 &  2.2\\
157P/Tritton                     &   2 &   2 & 2009-09-10 & 2009-11-07 & 17.2 & 18.5 &  1.8 &  2.2 &  1.0 &  1.2\\
158P/Kowal-LINEAR                &  17 &   7 & 2012-07-22 & 2012-07-29 & 18.8 & 19.4 &  4.6 &  4.6 &  4.1 &  4.2\\
160P/LINEAR                      &   7 &   2 & 2010-03-28 & 2012-07-18 & 18.7 & 19.2 &  2.1 &  5.3 &  1.4 &  4.3\\
162P/Siding Spring               &  16 &  10 & 2010-11-13 & 2012-03-21 & 18.8 & 20.5 &  2.7 &  4.7 &  3.0 &  3.8\\
163P/NEAT                        &   3 &   1 & 2011-11-03 & 2011-11-03 & 19.9 & 20.2 &  2.4 &  2.4 &  1.5 &  1.5\\
164P/Christensen                 &   4 &   3 & 2011-09-04 & 2011-09-08 & 17.9 & 18.8 &  1.9 &  1.9 &  2.6 &  2.6\\
167P/CINEOS                      &  17 &  15 & 2009-06-24 & 2010-10-29 & 20.7 & 21.6 & 13.9 & 14.5 & 13.0 & 14.1\\
169P/NEAT                        &   1 &   1 & 2009-07-07 & 2009-07-07 & 19.0 & 19.0 &  2.2 &  2.2 &  1.3 &  1.3\\
188P/LINEAR-Mueller              &   1 &   1 & 2010-02-19 & 2010-02-19 & 21.2 & 21.2 &  4.9 &  4.9 &  4.0 &  4.0\\
202P/Scotti                      &   3 &   1 & 2009-03-17 & 2009-03-17 & 19.5 & 19.8 &  2.5 &  2.5 &  2.5 &  2.5\\
203P/Korlevic                    &   3 &   2 & 2011-01-01 & 2011-01-08 & 17.6 & 18.4 &  3.6 &  3.6 &  2.9 &  3.0\\
213P/Van Ness                    &   2 &   2 & 2012-01-04 & 2012-01-05 & 17.2 & 17.3 &  2.5 &  2.5 &  2.6 &  2.6\\
215P/NEAT                        &  14 &   6 & 2011-11-02 & 2012-01-21 & 18.1 & 19.7 &  3.8 &  3.9 &  2.9 &  4.0\\
217P/LINEAR                      &  18 &  10 & 2009-06-26 & 2010-03-28 & 10.4 & 18.8 &  1.2 &  2.5 &  0.6 &  2.4\\
218P/LINEAR                      &   5 &   2 & 2009-05-25 & 2009-05-27 & 19.1 & 19.7 &  1.7 &  1.7 &  0.9 &  0.9\\
219P/LINEAR                      &  40 &  12 & 2010-08-13 & 2010-11-08 & 17.4 & 19.2 &  2.6 &  2.8 &  1.8 &  2.4\\
220P/McNaught                    &   3 &   1 & 2009-06-01 & 2009-06-01 & 19.9 & 20.5 &  2.3 &  2.3 &  1.4 &  1.4\\
221P/LINEAR                      &   7 &   6 & 2009-08-13 & 2009-09-20 & 20.4 & 21.2 &  2.5 &  2.6 &  1.6 &  1.7\\
223P/Skiff                       &   8 &   5 & 2010-08-13 & 2010-09-03 & 19.4 & 20.3 &  2.4 &  2.4 &  1.9 &  2.1\\
224P/LINEAR-NEAT                 &   1 &   1 & 2009-09-14 & 2009-09-14 & 21.4 & 21.4 &  2.3 &  2.3 &  1.3 &  1.3\\
225P/LINEAR                      &   2 &   2 & 2009-10-22 & 2009-10-22 & 20.2 & 20.6 &  1.5 &  1.5 &  0.9 &  0.9\\
226P/Pigott-LINEAR-Kowalski      &   2 &   2 & 2009-10-16 & 2009-10-16 & 19.3 & 19.6 &  2.3 &  2.3 &  2.2 &  2.2\\
228P/LINEAR                      &  12 &  10 & 2010-12-31 & 2012-03-05 & 18.0 & 20.4 &  3.5 &  3.5 &  2.6 &  3.3\\
229P/Gibbs                       &   4 &   2 & 2009-08-19 & 2009-08-23 & 19.7 & 20.2 &  2.4 &  2.4 &  2.1 &  2.1\\
22P/Kopff                        &   7 &   4 & 2009-06-26 & 2009-08-02 & 11.9 & 12.3 &  1.6 &  1.7 &  0.8 &  0.9\\
230P/LINEAR                      &   5 &   3 & 2009-12-03 & 2010-01-12 & 18.4 & 18.8 &  1.9 &  2.1 &  1.4 &  1.6\\
234P/LINEAR                      &   1 &   1 & 2009-12-15 & 2009-12-15 & 20.6 & 20.6 &  2.9 &  2.9 &  3.1 &  3.1\\
236P/LINEAR                      &  19 &  14 & 2010-06-17 & 2011-01-25 & 17.1 & 20.7 &  1.9 &  2.2 &  0.9 &  1.9\\
237P/LINEAR                      &  27 &  15 & 2010-07-05 & 2010-10-02 & 19.6 & 21.2 &  2.8 &  3.0 &  2.0 &  2.3\\
240P/NEAT                        &   7 &   5 & 2010-07-25 & 2010-12-08 & 14.5 & 16.6 &  2.1 &  2.2 &  1.3 &  2.6\\
241P/LINEAR                      &   8 &   8 & 2010-12-28 & 2011-02-01 & 17.4 & 18.4 &  2.4 &  2.6 &  1.6 &  1.7\\
\hline
\end{tabular}
\hfill{}
\end{table*}

\begin{table*}
 \renewcommand\thetable{3}
\caption{\emph{ --- Continued}}
\hfill{}
\begin{tabular}{lrrccrrrrrr}
\hline\\[-3ex]
name&obs.&nights&first date&last date& $V_\text{min}$& $V_\text{max}$& $r_\text{min}$& $r_\text{max}$& $\Delta_\text{min}$& $\Delta_\text{max}$\\
\hline
242P/Spahr                       &  16 &  13 & 2010-08-15 & 2011-08-28 & 19.3 & 21.1 &  4.1 &  4.8 &  3.7 &  4.4\\
243P/NEAT                        &   5 &   3 & 2011-11-03 & 2011-11-22 & 20.2 & 20.9 &  2.9 &  3.0 &  2.1 &  2.2\\
244P/Scotti                      &  29 &  17 & 2010-09-10 & 2011-01-06 & 19.3 & 20.3 &  4.2 &  4.3 &  3.3 &  3.9\\
245P/WISE                        &   9 &   6 & 2010-07-25 & 2010-09-11 & 19.1 & 20.4 &  2.5 &  2.7 &  1.7 &  1.7\\
246P/NEAT                        &  14 &   7 & 2011-02-13 & 2011-11-23 & 16.5 & 19.1 &  3.6 &  4.3 &  3.4 &  4.1\\
247P/LINEAR                      &  12 &   7 & 2010-10-08 & 2010-11-12 & 17.1 & 20.2 &  1.6 &  1.8 &  0.7 &  1.1\\
248P/Gibbs                       &  25 &  15 & 2010-09-18 & 2010-12-11 & 18.2 & 19.8 &  2.2 &  2.5 &  1.4 &  1.7\\
250P/Larson                      &   3 &   2 & 2010-11-03 & 2010-11-04 & 20.4 & 20.8 &  2.2 &  2.2 &  2.1 &  2.2\\
253P/PANSTARRS                   &   4 &   3 & 2011-11-02 & 2011-12-08 & 16.9 & 18.0 &  2.0 &  2.0 &  1.3 &  1.6\\
254P/McNaught                    &   1 &   1 & 2011-11-03 & 2011-11-03 & 17.8 & 17.8 &  3.7 &  3.7 &  3.0 &  3.0\\
260P/McNaught                    &   9 &   3 & 2012-07-27 & 2012-07-30 & 14.2 & 14.7 &  1.6 &  1.6 &  0.9 &  0.9\\
261P/Larson                      &  18 &   9 & 2012-06-25 & 2012-07-06 & 19.2 & 20.3 &  2.3 &  2.3 &  1.7 &  1.8\\
279P/La Sagra                    &   6 &   3 & 2009-07-20 & 2009-08-02 & 20.1 & 21.4 &  2.2 &  2.2 &  1.3 &  1.4\\
P/2006 VW$_{139}$                &   5 &   3 & 2011-09-27 & 2011-12-21 & 19.0 & 19.6 &  2.5 &  2.6 &  1.5 &  1.9\\
P/2009 O3 (Hill)                  &   7 &   5 & 2009-09-20 & 2009-11-07 & 17.5 & 18.5 &  2.7 &  2.9 &  1.8 &  2.0\\
P/2009 Q1 (Hill)                &   5 &   3 & 2009-08-01 & 2010-12-31 & 18.4 & 19.9 &  2.8 &  4.5 &  2.0 &  3.7\\
P/2009 Q4 (Boattini)              &   6 &   4 & 2009-12-16 & 2010-03-16 & 13.4 & 17.8 &  1.4 &  1.8 &  0.6 &  0.9\\
P/2009 Q5 (McNaught)             &   2 &   1 & 2009-08-21 & 2009-08-21 & 17.0 & 17.1 &  2.9 &  2.9 &  2.2 &  2.2\\
P/2009 SK$_{280}$ (Spacewatch-Hill)     &   5 &   3 & 2009-10-23 & 2009-11-09 & 19.8 & 20.4 &  4.2 &  4.2 &  3.2 &  3.3\\
P/2009 T2 (La Sagra)               &   8 &   6 & 2009-08-24 & 2010-03-13 & 16.5 & 20.5 &  1.8 &  2.3 &  1.1 &  1.9\\
P/2009 WX$_{51}$ (Catalina)             &   6 &   1 & 2009-12-17 & 2009-12-17 & 17.4 & 18.0 &  1.1 &  1.1 &  0.2 &  0.2\\
P/2010 A3 (Hill)                   &   3 &   2 & 2009-09-13 & 2010-03-25 & 16.1 & 21.3 &  1.6 &  2.7 &  1.8 &  1.9\\
P/2010 A5 (LINEAR)                 &   4 &   3 & 2010-01-12 & 2010-02-24 & 16.3 & 17.4 &  1.8 &  2.0 &  1.2 &  1.8\\
P/2010 B2 (WISE)                   &   1 &   1 & 2010-02-23 & 2010-02-23 & 20.0 & 20.0 &  1.7 &  1.7 &  1.1 &  1.1\\
P/2010 D2 (WISE)                   &   1 &   1 & 2010-03-17 & 2010-03-17 & 19.9 & 19.9 &  3.7 &  3.7 &  3.6 &  3.6\\
P/2010 E2 (Jarnac)                 &   5 &   4 & 2010-06-08 & 2010-06-28 & 19.1 & 20.2 &  2.5 &  2.5 &  2.0 &  2.3\\
P/2010 H2 (Vales)                  &  11 &   6 & 2010-04-16 & 2010-06-01 & 11.8 & 15.1 &  3.1 &  3.1 &  2.1 &  2.4\\
P/2010 H5 (Scotti)                 &  18 &  10 & 2010-05-30 & 2010-06-17 & 20.4 & 21.2 &  6.0 &  6.0 &  5.4 &  5.6\\
P/2010 N1 (WISE)                   &   2 &   2 & 2010-03-12 & 2010-03-15 & 20.9 & 21.1 &  2.1 &  2.1 &  1.2 &  1.2\\
P/2010 P4 (WISE)                   &   3 &   3 & 2010-09-15 & 2010-10-04 & 20.1 & 20.8 &  2.0 &  2.0 &  1.2 &  1.3\\
P/2010 R2 (La Sagra)               &  34 &  21 & 2010-07-06 & 2010-10-29 & 18.2 & 20.1 &  2.6 &  2.7 &  1.7 &  2.1\\
P/2010 T2 (PANSTARRS)              &   7 &   5 & 2010-09-05 & 2010-09-15 & 19.9 & 21.4 &  4.0 &  4.0 &  3.1 &  3.2\\
P/2010 TO$_{20}$ (LINEAR-Grauer)        &   1 &   1 & 2010-08-24 & 2010-08-24 & 18.7 & 18.7 &  5.3 &  5.3 &  4.4 &  4.4\\
P/2010 U1 (Boattini)               &  62 &  33 & 2009-06-25 & 2010-11-13 & 19.1 & 21.4 &  4.9 &  5.0 &  4.0 &  4.9\\
P/2010 U2 (Hill)                   &  34 &  19 & 2010-09-07 & 2010-12-13 & 17.6 & 19.9 &  2.6 &  2.6 &  1.6 &  1.9\\
P/2010 UH$_{55}$ (Spacewatch)           &  24 &  10 & 2010-10-15 & 2011-10-17 & 18.7 & 20.3 &  3.0 &  3.2 &  2.1 &  3.6\\
P/2010 WK (LINEAR)                 &   9 &   5 & 2010-08-14 & 2010-09-22 & 18.7 & 20.4 &  1.8 &  1.9 &  1.2 &  1.6\\
P/2011 C2 (Gibbs)                  &  28 &  19 & 2010-12-02 & 2012-02-01 & 19.7 & 21.1 &  5.4 &  5.6 &  4.7 &  5.3\\
P/2011 JB$_{15}$ (Spacewatch-Boattini)  &   2 &   2 & 2010-06-06 & 2010-06-09 & 20.7 & 21.1 &  5.6 &  5.6 &  5.0 &  5.0\\
P/2011 NO$_1$ (Elenin)                &   1 &   1 & 2011-07-30 & 2011-07-30 & 19.7 & 19.7 &  2.6 &  2.6 &  1.6 &  1.6\\
P/2011 P1 (McNaught)               &   4 &   2 & 2011-09-08 & 2011-09-20 & 18.7 & 19.4 &  5.3 &  5.3 &  4.6 &  4.7\\
P/2011 Q3 (McNaught)               &  38 &  20 & 2011-07-23 & 2011-11-30 & 18.5 & 20.5 &  2.4 &  2.5 &  1.4 &  2.0\\
P/2011 R3 (Novichonok)             &   5 &   3 & 2011-10-08 & 2011-10-10 & 18.3 & 18.6 &  3.7 &  3.7 &  2.7 &  2.7\\
P/2011 U1 (PANSTARRS)             &  18 &   7 & 2011-11-24 & 2012-01-18 & 19.1 & 20.9 &  2.6 &  2.8 &  1.8 &  1.9\\
P/2011 VJ$_5$ (Lemmon)                &   5 &   3 & 2012-02-04 & 2012-03-25 & 18.5 & 19.4 &  1.6 &  1.9 &  0.9 &  0.9\\
P/2012 B1 (PANSTARRS)              &   9 &   6 & 2011-12-11 & 2012-01-04 & 19.2 & 19.9 &  4.9 &  4.9 &  4.0 &  4.3\\
C/2002 VQ$_{94}$ (LINEAR) &   3 &   2 & 2009-06-09 & 2009-07-06 & 18.6 & 18.7 & 10.2 & 10.3 &  9.5 & 10.0\\
C/2005 EL$_{173}$ (LONEOS) &   2 &   1 & 2009-07-28 & 2009-07-28 & 19.5 & 20.4 &  8.0 &  8.0 &  7.3 &  7.3\\
C/2005 L3 (McNaught) &  47 &  19 & 2009-05-12 & 2012-03-16 & 14.0 & 19.4 &  6.6 & 11.6 &  6.0 & 10.8\\
C/2006 OF$_2$ (Broughton) &  89 &   5 & 2010-01-25 & 2010-02-16 & 16.0 & 16.6 &  5.5 &  5.7 &  4.6 &  4.7\\
C/2006 Q1 (McNaught) &  11 &   6 & 2009-05-13 & 2009-08-16 & 14.0 & 15.1 &  4.2 &  4.8 &  3.5 &  4.8\\
C/2006 S3 (LONEOS) &  37 &  25 & 2009-06-25 & 2010-09-12 & 15.4 & 17.5 &  6.7 &  8.9 &  5.8 &  8.6\\
C/2006 U6 (Spacewatch) &   4 &   2 & 2009-03-25 & 2010-03-16 & 16.2 & 19.7 &  3.9 &  6.6 &  3.0 &  5.7\\
C/2007 D1 (LINEAR) &   6 &   3 & 2010-03-12 & 2011-03-16 & 17.8 & 18.8 & 10.5 & 11.7 &  9.6 & 10.9\\
C/2007 G1 (LINEAR) &   7 &   5 & 2010-12-28 & 2011-01-23 & 19.0 & 20.1 &  7.5 &  7.7 &  6.7 &  7.1\\
C/2007 M1 (McNaught) &  14 &   8 & 2009-07-04 & 2010-03-17 & 18.6 & 20.5 &  7.7 &  8.3 &  7.1 &  8.0\\
C/2007 N3 (Lulin) &   2 &   1 & 2009-12-27 & 2009-12-27 & 16.3 & 16.5 &  4.6 &  4.6 &  3.6 &  3.6\\
C/2007 Q3 (Siding Spring) &  33 &  20 & 2009-11-03 & 2010-07-23 & 11.1 & 14.6 &  2.3 &  3.8 &  2.2 &  3.9\\
C/2007 T5 (Gibbs) &   7 &   5 & 2009-05-16 & 2009-06-29 & 19.9 & 20.5 &  5.0 &  5.2 &  4.5 &  5.1\\
C/2007 U1 (LINEAR) &  21 &  13 & 2009-06-24 & 2009-09-07 & 16.7 & 17.8 &  4.4 &  4.9 &  3.9 &  4.4\\
C/2007 VO$_{53}$ (Spacewatch) &  26 &  11 & 2011-06-24 & 2012-06-27 & 17.8 & 20.7 &  5.8 &  7.6 &  5.5 &  7.0\\
C/2008 FK$_{75}$ (Lemmon-Siding Spring) &  19 &  11 & 2009-06-28 & 2010-09-28 & 15.3 & 16.7 &  4.5 &  5.8 &  4.1 &  5.2\\
C/2008 N1 (Holmes) &  13 &   7 & 2009-07-21 & 2010-06-06 & 16.3 & 18.3 &  2.8 &  3.8 &  2.6 &  3.6\\
C/2008 P1 (Garradd) &   4 &   2 & 2009-08-23 & 2009-09-14 & 15.5 & 15.8 &  3.9 &  3.9 &  3.0 &  3.2\\
\hline
\end{tabular}
\hfill{}
\end{table*}

\begin{table*}
 \renewcommand\thetable{3}
\caption{\emph{ --- Continued}}
\hfill{}
\begin{tabular}{lrrccrrrrrr}
\hline\\[-3ex]
name&obs.&nights&first date&last date& $V_\text{min}$& $V_\text{max}$& $r_\text{min}$& $r_\text{max}$& $\Delta_\text{min}$& $\Delta_\text{max}$\\
\hline
C/2008 Q1 (Maticic) &  15 &   7 & 2009-05-13 & 2011-02-22 & 16.1 & 18.8 &  3.2 &  7.5 &  2.6 &  6.7\\
C/2008 Q3 (Garradd) &   2 &   1 & 2010-03-26 & 2010-03-26 & 17.6 & 17.8 &  3.7 &  3.7 &  3.1 &  3.1\\
C/2008 S3 (Boattini) &  33 &  12 & 2010-09-29 & 2010-11-06 & 17.4 & 18.4 &  8.1 &  8.2 &  7.2 &  7.3\\
C/2009 F1 (Larson) &   2 &   1 & 2009-03-27 & 2009-03-27 & 18.7 & 18.8 &  2.1 &  2.1 &  1.2 &  1.2\\
C/2009 F2 (McNaught) &   2 &   2 & 2012-06-26 & 2012-06-28 & 20.8 & 20.8 &  8.7 &  8.7 &  8.1 &  8.1\\
C/2009 K2 (Catalina) &  19 &  10 & 2009-05-08 & 2009-08-24 & 17.8 & 19.8 &  3.6 &  4.1 &  3.6 &  3.9\\
C/2009 K5 (McNaught) &   3 &   3 & 2010-09-27 & 2010-11-12 & 13.7 & 14.4 &  2.5 &  3.0 &  2.3 &  2.6\\
C/2009 O2 (Catalina) &   4 &   2 & 2009-06-29 & 2009-07-21 & 19.8 & 21.3 &  3.7 &  4.0 &  2.7 &  3.2\\
C/2009 P1 (Garradd) &   7 &   4 & 2011-07-21 & 2012-02-02 &  8.6 &  9.5 &  1.6 &  2.6 &  1.4 &  1.7\\
C/2009 P2 (Boattini) &  44 &  26 & 2009-08-13 & 2010-09-14 & 18.5 & 19.8 &  6.6 &  6.7 &  5.7 &  6.8\\
C/2009 T3 (LINEAR) &   1 &   1 & 2010-06-03 & 2010-06-03 & 18.9 & 18.9 &  2.8 &  2.8 &  2.9 &  2.9\\
C/2009 U3 (Hill) &  20 &   4 & 2010-01-17 & 2010-05-04 & 16.2 & 16.7 &  1.5 &  1.7 &  1.3 &  1.4\\
C/2009 U5 (Grauer) &   5 &   4 & 2010-12-08 & 2011-01-12 & 20.2 & 21.1 &  6.2 &  6.3 &  5.7 &  6.1\\
C/2009 UG$_{89}$ (Lemmon) &  61 &  32 & 2011-04-27 & 2012-04-29 & 17.0 & 20.2 &  4.1 &  5.7 &  3.6 &  5.2\\
C/2009 Y1 (Catalina) &  14 &   8 & 2009-12-30 & 2011-09-28 & 15.2 & 19.4 &  3.5 &  4.7 &  2.7 &  4.2\\
C/2010 B1 (Cardinal) &   5 &   3 & 2010-01-11 & 2010-01-25 & 17.8 & 18.0 &  4.7 &  4.7 &  4.0 &  4.1\\
C/2010 D4 (WISE) &  32 &  21 & 2009-05-18 & 2010-09-18 & 19.8 & 21.3 &  7.2 &  7.8 &  6.5 &  8.2\\
C/2010 DG$_{56}$ (WISE) &  13 &   9 & 2010-07-26 & 2010-09-11 & 18.1 & 20.0 &  1.9 &  2.2 &  1.1 &  1.6\\
C/2010 E5 (Scotti) &   3 &   2 & 2010-03-19 & 2010-03-25 & 19.8 & 19.9 &  4.0 &  4.0 &  3.0 &  3.0\\
C/2010 F1 (Boattini) &  11 &   9 & 2009-11-09 & 2010-01-17 & 18.5 & 19.5 &  3.6 &  3.6 &  3.0 &  3.7\\
C/2010 G2 (Hill) &  10 &   7 & 2010-06-23 & 2012-01-15 & 12.4 & 18.9 &  2.5 &  5.0 &  2.1 &  4.5\\
C/2010 G3 (WISE) &  37 &  24 & 2009-10-03 & 2011-06-26 & 18.6 & 20.4 &  4.9 &  5.9 &  4.7 &  6.3\\
C/2010 J1 (Boattini) &   4 &   2 & 2010-06-13 & 2010-06-24 & 18.1 & 19.2 &  2.3 &  2.4 &  1.7 &  2.0\\
C/2010 J2 (McNaught) &   6 &   4 & 2010-06-27 & 2011-06-10 & 16.9 & 20.0 &  3.4 &  4.8 &  2.6 &  4.2\\
C/2010 L3 (Catalina) &  53 &  34 & 2009-08-03 & 2012-07-16 & 18.8 & 20.9 &  9.9 & 10.4 &  9.3 & 10.3\\
C/2010 R1 (LINEAR) &  27 &  11 & 2012-06-01 & 2012-06-27 & 16.9 & 17.4 &  5.6 &  5.6 &  4.8 &  5.1\\
C/2010 S1 (LINEAR) &   1 &   1 & 2010-02-18 & 2010-02-18 & 20.3 & 20.3 &  9.9 &  9.9 &  9.8 &  9.8\\
C/2010 U3 (Boattini) &   6 &   5 & 2010-10-17 & 2011-09-04 & 20.0 & 20.6 & 17.1 & 18.4 & 16.6 & 17.5\\
C/2010 X1 (Elenin) &  23 &  18 & 2011-01-06 & 2011-02-22 & 17.6 & 19.4 &  3.3 &  3.9 &  2.4 &  3.6\\
C/2011 A3 (Gibbs) &  22 &   9 & 2011-03-04 & 2011-04-15 & 16.6 & 17.5 &  3.5 &  3.8 &  2.7 &  3.1\\
C/2011 C1 (McNaught) &   1 &   1 & 2011-08-25 & 2011-08-25 & 19.7 & 19.7 &  2.3 &  2.3 &  1.7 &  1.7\\
C/2011 C3 (Gibbs) &   1 &   1 & 2011-02-11 & 2011-02-11 & 20.4 & 20.4 &  1.7 &  1.7 &  1.3 &  1.3\\
C/2011 F1 (LINEAR) &  27 &  18 & 2010-10-12 & 2012-05-24 & 13.4 & 19.7 &  3.3 &  8.2 &  2.9 &  8.6\\
C/2011 G1 (McNaught) &   5 &   2 & 2011-11-08 & 2012-01-27 & 17.3 & 17.5 &  2.2 &  2.6 &  1.8 &  2.5\\
C/2011 J3 (LINEAR) &   1 &   1 & 2011-06-23 & 2011-06-23 & 19.0 & 19.0 &  2.4 &  2.4 &  2.1 &  2.1\\
C/2011 L3 (McNaught) &  18 &  10 & 2011-07-15 & 2011-10-01 & 14.9 & 17.6 &  1.9 &  2.0 &  1.0 &  1.9\\
C/2011 M1 (LINEAR) &   2 &   1 & 2011-07-04 & 2011-07-04 & 15.7 & 16.0 &  1.4 &  1.4 &  1.4 &  1.4\\
C/2011 P2 (PANSTARRS) &  11 &   8 & 2011-06-10 & 2011-08-19 & 19.4 & 20.1 &  6.3 &  6.3 &  5.3 &  5.5\\
C/2011 Q4 (SWAN) &   4 &   2 & 2012-02-25 & 2012-02-26 & 20.3 & 20.7 &  2.5 &  2.5 &  1.8 &  1.8\\
C/2011 R1 (McNaught) &   9 &   6 & 2010-10-04 & 2010-12-13 & 19.5 & 20.7 &  7.0 &  7.5 &  6.3 &  6.6\\
C/2012 A1 (PANSTARRS) &   4 &   4 & 2010-10-31 & 2011-01-01 & 20.4 & 21.2 & 10.0 & 10.2 &  9.3 & 10.3\\
C/2012 A2 (LINEAR) &   3 &   2 & 2011-04-07 & 2011-11-18 & 18.9 & 20.9 &  4.7 &  6.1 &  5.1 &  5.2\\
C/2012 CH$_{17}$ (MOSS) &   3 &   2 & 2012-01-04 & 2012-01-06 & 19.2 & 20.1 &  3.7 &  3.7 &  3.1 &  3.2\\
C/2012 E1 (Hill) &  54 &  26 & 2011-06-10 & 2012-05-29 & 19.4 & 20.5 &  7.5 &  7.8 &  6.7 &  7.4\\
C/2012 E3 (PANSTARRS) &   1 &   1 & 2012-06-09 & 2012-06-09 & 20.5 & 20.5 &  5.1 &  5.1 &  4.8 &  4.8\\
C/2012 Q1 (Kowalski) &   2 &   1 & 2011-10-01 & 2011-10-01 & 19.9 & 20.1 &  9.5 &  9.5 &  8.9 &  8.9\\
\hline
\end{tabular}
\hfill{}
\end{table*}

Of the asteroids that were observed by the \emph{Sloan Digital Sky Survey} (SDSS; \citealp{yor00}) during its 1998--2009 imaging phase, 142,774 known objects have $g$, $r$ and $i$ photometry with errors of less than 0.2 mag in all three bands, and more than half (72,556) of these objects were also observed by PTF. These data come from the SDSS Moving Object Catalog 4th release \citep{par08}, which includes data through March 2007, supplemented with more recent SDSS moving object data from 2008--2009 (B. Sesar, personal communication). The principal component color $a^*=0.89(g-r)+0.45(r-i)-0.57$ is useful for broad (C-type vs. S-type) taxonomic classification (\citealp{ive02}; \citealp{par08}). Relative to this known-color sample, PTF detected 49\% of the carbonaceous-colored ($a^*<0$) and 52\% of the stony-colored ($a^*>0$) asteroids (Figure 3, bottom right).

Of the 27,326 objects that were observed by \emph{both} WISE and SDSS (satisfying the measurement error constraints mentioned above), 16,955 (62\%) of these were also observed by PTF. A total of 624 of these WISE+SDSS objects were observed at least 10 times on at least one night in PTF, whence rotation curves can be estimated, while 625 of these WISE+SDSS objects have PTF observations in five or more phase-angle bins of width 3$^{\circ}$, including opposition (0$^{\circ}$ --3$^{\circ}$), whence phase functions can be estimated.

\begin{figure}
\centering
\includegraphics[scale=0.13]{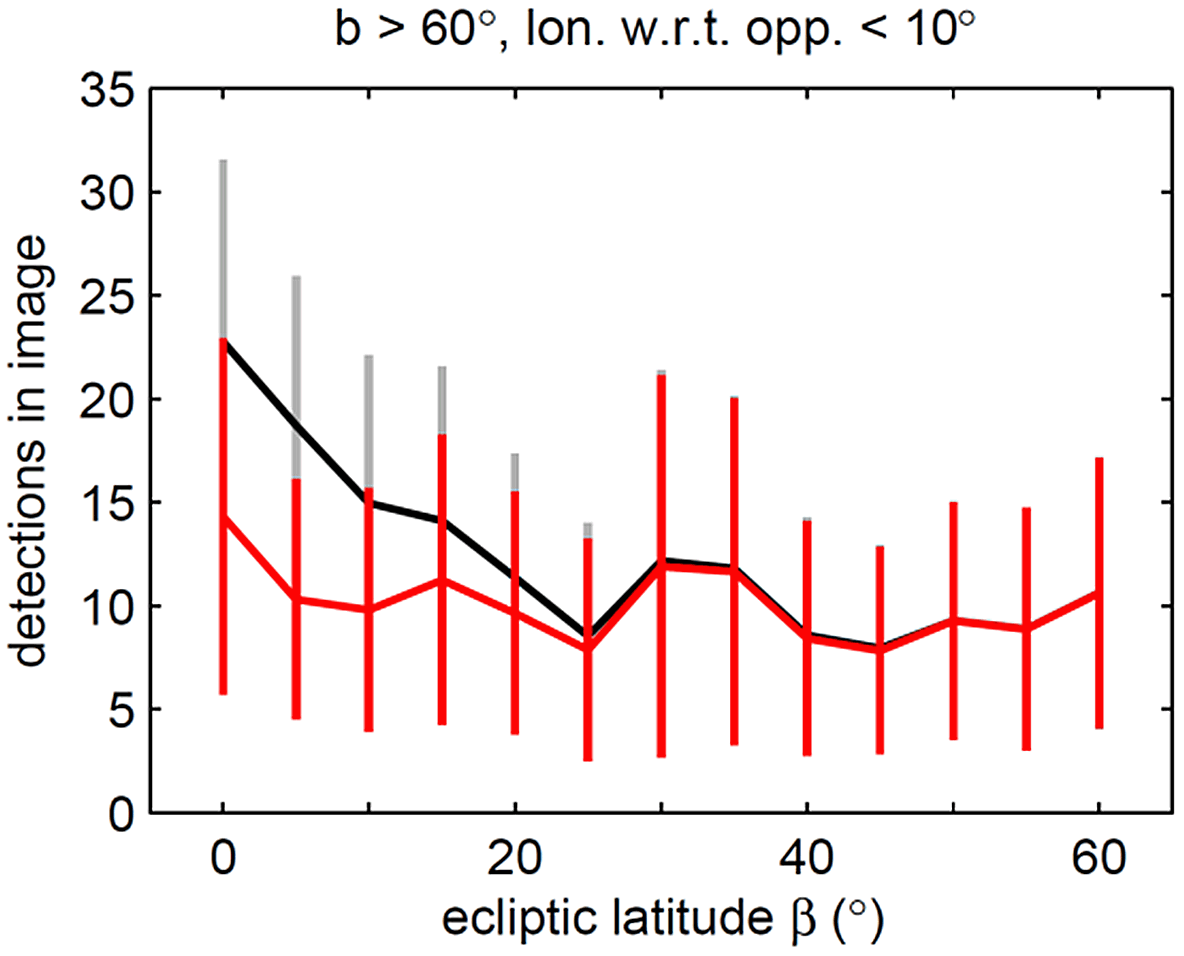}
\caption{As in Figure 2, the vertical bars show the scatter (standard deviation) and the connected points are the mean values for the bins. Now added in red is the distribution of transients after exclusion of the $\sim$2 million known-object detections. The original distribution is included, in black, for comparison.}
\end{figure}

\section{Unknown-object extraction}

Exclusion of the $\sim$2 million known-object detections leaves $\sim$18 million transients remaining in our list of moving-object candidates. Figure 5 shows that the ecliptic distribution of transients per image has flattened out substantially. However, ignoring the scatter, the mean number of transients in the leftmost (lowest ecliptic latitude) bin remains the highest by more than two detections per image, suggesting the presence of significant unknown ({\it i.e.}, undiscovered) small bodies in the data.

\subsection{Previous and ongoing PTF small-body discovery work}

In \cite{pol12}, a pilot study of rotation curve analysis and new-object discovery was undertaken using a few nights of $\sim$$20\deg^2$  high cadence ($\sim$20-minute-spaced) PTF data obtained in February 2010 at low ecliptic latitude ($|\beta|<2.5^{\circ}$). Using an original moving-object detection algorithm, they extracted 684 asteroids; of those which received provisional designations, three still qualify as PTF discoveries as of March 2013 (2010 CU$_{247}$, 2010 CL$_{249}$ and 2010 CN$_{249}$). Though highly efficient on high cadence data, their tracklet-finding algorithm's limitations ({\it e.g.}, single-night, single-CCD) renders it inapplicable to the vast majority of regular- (hour-to-days-) cadence PTF data. 

A popular solution to this problem was already mentioned in Section 3, namely the use of kd-trees. A recently successful such kd-tree-based, detection-couplet-matching MOPS was used on the WISE data \citep{dai10}. The WISE MOPS successfully extracted $\sim$2 million observations of $\sim$158,000 moving objects from the WISE data, including $\sim$34,000 new objects. A modified version of the WISE MOPS is under development for PTF at IPAC. As with the WISE MOPS, a key intent is the discovery of near-Earth objects, hence the PTF MOPS will need to accommodate relatively fast apparent motions (at least an order of magnitude faster than main-belt speeds). This poses considerable challenges, because PTF's cadences and false-positive detection rates are less accommodating than those of the space-based WISE survey. Though far from complete, the prototype PTF MOPS has successfully demonstrated that it can find tracklets spanning multiple nights and multiple fields of view, including at least two near-Earth objects, one of which was unknown (J. Bauer, personal communication).

As the PTF MOPS is still in development, for the purposes of this work we implement an original moving-object detection algorithm and run it on our residual $\sim$18 million-transient sample. The reader who wishes to skip over the details of the discovery algorithm should now go to Section 4.3. 

\subsection{A custom discovery algorithm for main-belt objects}

Because our intention is solely to supplement the main-belt comet search, we restrict apparent motions to those typical of main-belt objects (thereby easing the computational burden, but excluding faster NEOs and slower TNOs). This on-sky motion range is taken to be between 0.1 and 1.0 arcsec/minute.

Analysis of the known-object sample (Figure 6) shows that about half of all consecutive-observation pairs occur over a less than 12-hour (same night) interval, with a sharp peak at the one-hour spacing. Of the remaining (multi-night) consecutive-observation pairs, roughly half span less than 48 hours. Given these statistics, we prescribe 2 days as our maximum allowable timespan (between first and last observation) for a minimum three-point tracklet. As will be explained, multiple primary tracklets can be merged to produce a secondary tracklet greater than 2 days in total length, but the interval between any two consecutive points in the secondary tracklet still will not exceed 2 days. An imposed minimum time of 10 minutes between consecutive tracklet points ensures that the object has moved at least one arcsecond (for the minimum allowed speed), such that stationary transients ({\it e.g.}, hostless-supernovae) are excluded.

\begin{figure}
\centering
\includegraphics[scale=0.19]{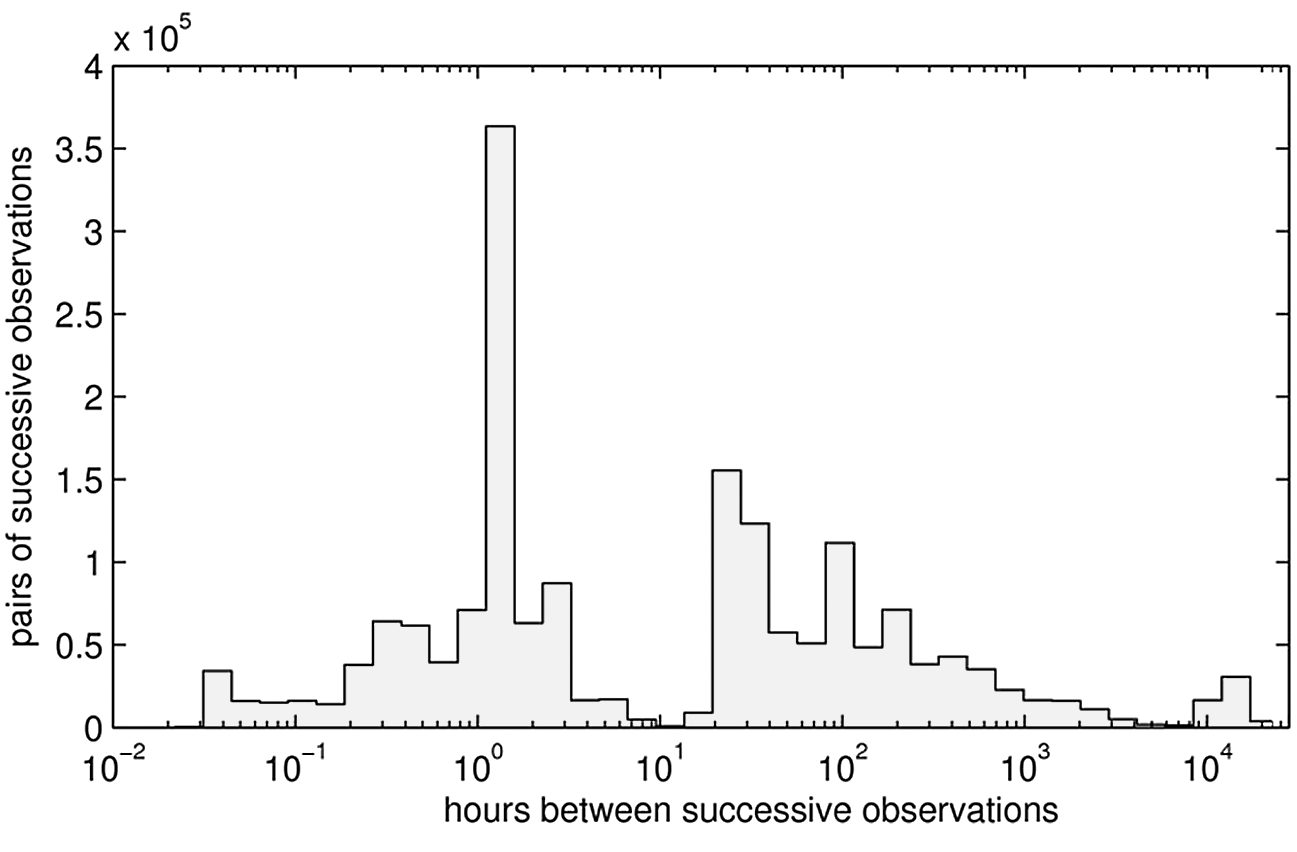}
\caption{Time interval between consecutive observations of known objects in PTF. This distribution justifies the 48-hour upper limit we impose for tracklet finding, which also was chosen for computational expediency.}
\end{figure}

Having specified time and velocity limits, the problem reduces to searching a double-cone-shaped volume, in three-dimensional time-plus-sky space surrounding each transient, to find sufficiently collinear past and future points. We modified our kd-tree implementation from Section 3 for this purpose. In particular, because PTF data were not collected on every consecutive night of the 41-months (due to weather, scheduling, etc.), the two-day upper limit we impose makes node-splitting along two-day (minimum) gaps in the data more natural and practical than simply splitting at median times, as was done in Section 3.1 and as is done generally for kd-trees.

An illustration of the tracklet-finding scheme (simplified to one spatial dimension) appears in Figure 7. For each transient, the kd-tree is used to rapidly find all other transients within its surrounding double-cone. Then, for every candidate past-plus-future pair of points, the two components of velocity and the distance residual of the middle transient from the candidate pair's predicted location (at the middle transient's epoch) are computed. Candidate past-plus-future pairs are then automatically discarded on the basis of the middle transient's distance residual with respect to them. To accommodate a limited amount of constant curvature, we use an adaptive criterion that is least stringent when the middle transient lies exactly at the midpoint between the past and future points, and becomes linearly more stringent as the middle transient nears one endpoint (approaching zero-tolerance at an endpoint). For candidate pairs spanning a single night or less, the maximum allowed middle-point residual is $1''$, while multi-night candidate pairs are allowed up to a $10''$ offset at the midpoint. 

\begin{figure}
\centering
\includegraphics[scale=0.16]{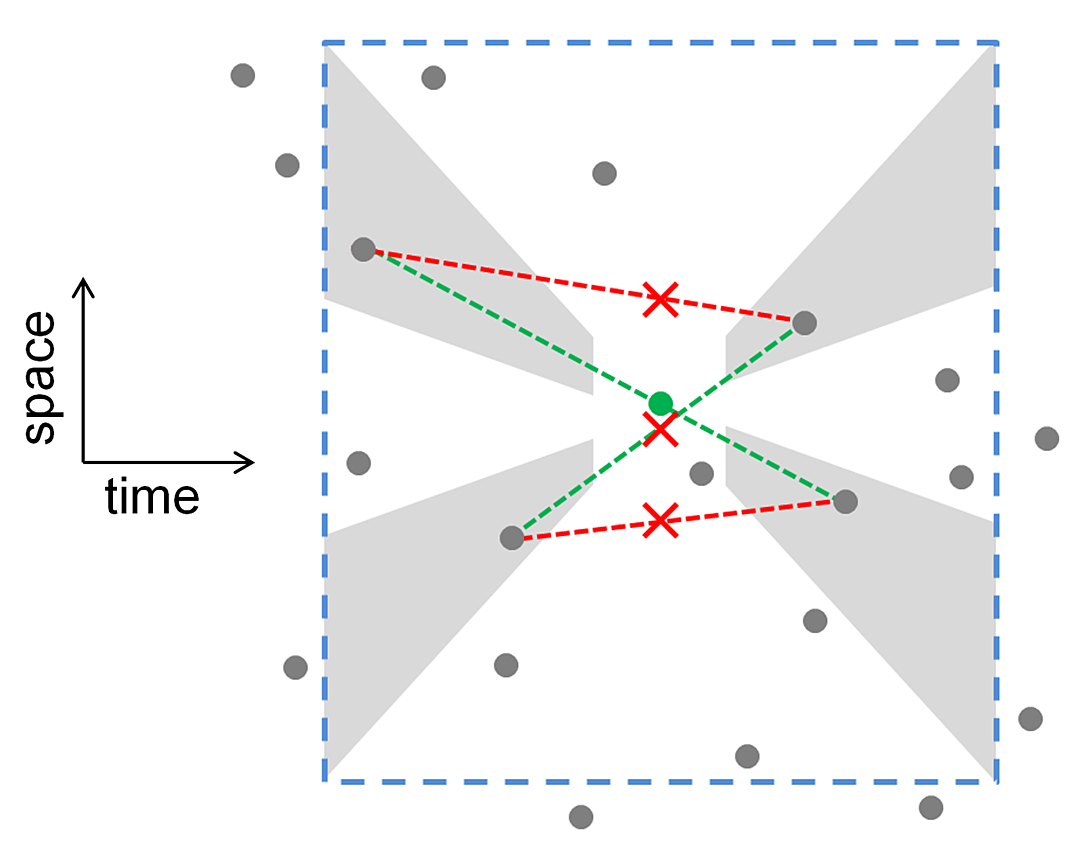}
\caption{Schematic of our tracklet-finding algorithm, with only one spatial dimension (rather than the actual two) for clarity. Given all transients (gray dots), a kd-tree search rapidly finds those that are “nearby” in spacetime (within the blue dashed box, of 48-hour full-width) to the central, target transient (green dot). Minimal- and maximal-velocity bounds then define a subset of these (all dots lying in the gray shaded regions, in this case four). All possible past + future pairs are considered (in this case, four possible pairs). Pairs whose predicted midpoint position is sufficiently far from the target transient are immediately rejected (red dashed lines). Pairs with a sufficiently small residual (green dashed lines) then are binned in velocity space, and the velocity bin containing the most pairs is chosen. In this example, however, both non-empty velocity bins have only one pair, in which case the pair with the smallest midpoint residual is chosen.}
\end{figure}

All remaining candidate past-plus-future pairs are then binned in two dimensions based on their two velocity components (R.A. and Dec. rates). Since the maximum allowed speed is 1 arcsec/minute, bins of 0.05 arcsec/minute between $\pm$1 arcsec/minute are used. If any single bin contains more candidate pairs than any other bin, all transients in all pairs in that bin, plus the middle transient, are automatically assigned a unique tracklet label. If more than one bin has the maximal number of pairs, the pair with the smallest midpoint residual is used. If any of these transients already has a tracklet label, all the others are instead assigned that existing label.

Following this stage in the new-object discovery process, all tracklets found are screened rapidly by eye to eliminate false positives. The remaining tracklets are assigned a preliminary orbital solution using the orbit-fitting software \emph{Find\_Orb}\footnote{The batch (non-interactive) Linux version of \emph{Find\_Orb} tries combinations of the V\"ais\"al\"a and Gauss orbit-determination methods on subsets of each tracklet in an attempt to converge on an orbit solution with minimized errors. For more information, see \href{http://www.projectpluto.com/find\_orb.htm}{\color{blue}{http://www.projectpluto.com/find\_orb.htm}}.} in batch mode. All orbital solutions are then used to re-search the transient data set for missed observations which could further refine the object's orbit. Because linear position extrapolation is replaced at this point by full orbital-solution-based ephemerides, the merging of tracklets across gaps in time longer than 48 hours is attempted in this last step.

\subsection{Summary of objects discovered}

We found 626 new objects which had a sufficient number of observations (at least two per night on at least two nights) to merit submission to the Minor Planet Center (MPC), whereupon they were assigned provisional designations. Four new comets were among the objects found by this moving-object search: 2009 KF$_{37}$, 2010 LN$_{135}$, 2012 KA$_{51}$, and C/2012 LP$_{26}$ (see Table 4 in Section 6 for details). The first is a Jupiter-family comet and the latter three are long-period comets. As of March 2013, the first three still bear provisional asteroidal designations assigned by the MPC's automated procedures; the fourth, C/2012 LP$_{26}$ (Palomar), was given its official cometary designation after follow-up observations were made in Februrary 2013 \citep{was13}. The cometary nature of these objects was initially noted on the basis of their orbital elements; an independent confirmation on the basis their measured extendedness appears in Section 6.

\begin{figure}
\centering
\includegraphics[scale=0.125]{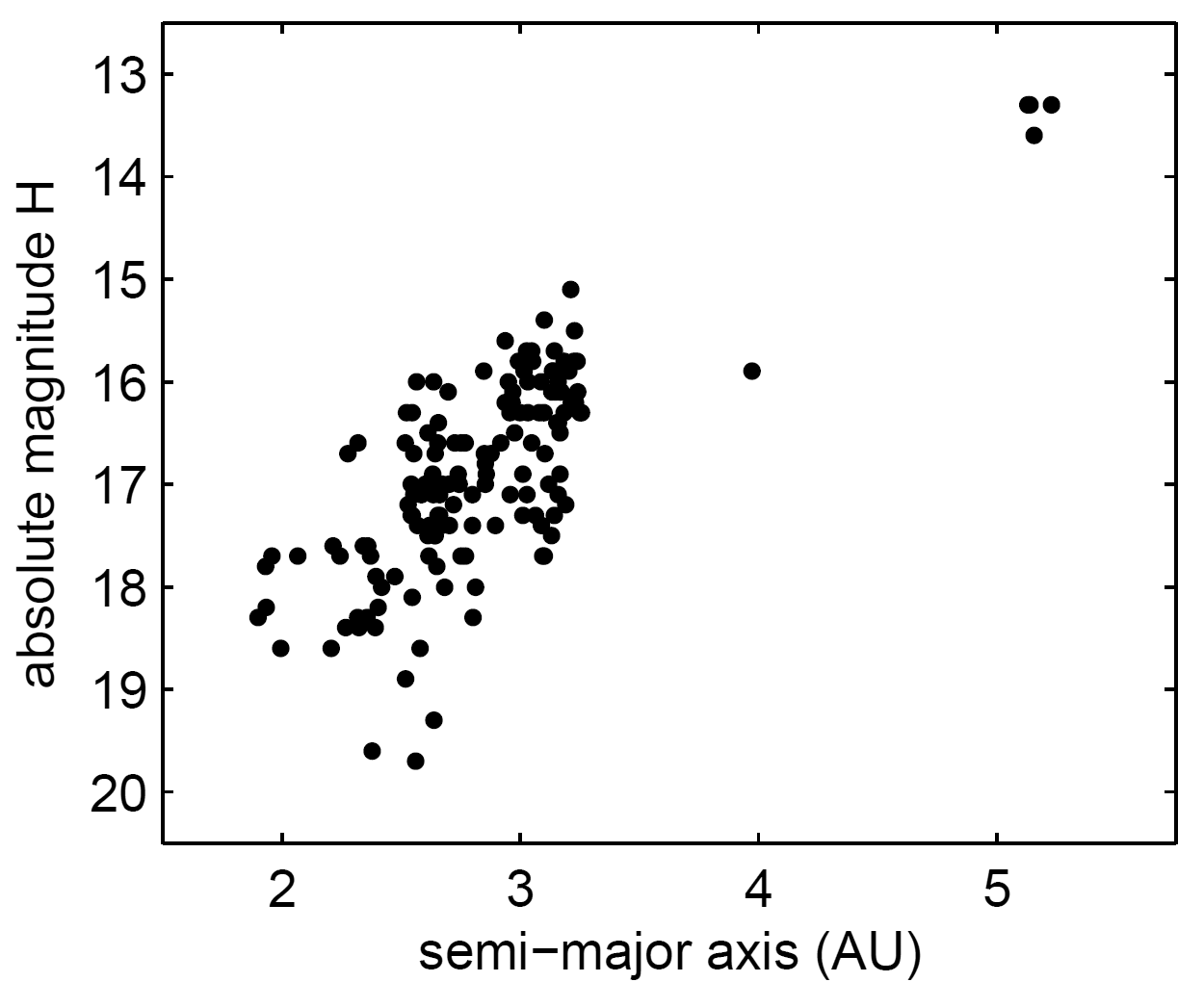}
\caption{Of 622 asteroids discovered in multi-night PTF data through July 2012, the MPC has provided orbital solutions for 470 of these. Assuming typical albedos, the smallest ($H\sim19.5$)  objects correspond to $\sim$0.5-km diameters, while the $H\sim13$ Trojans correspond to $\sim$10-km diameters.}
\end{figure}

\section{Extended-object analysis: Approach}

\subsection{Definition of the extendedness parameter $\mu$}

To quantify the extendedness of a given small-body observation, we use the ratio of the object's total flux, within a flexible elliptical aperture (Kron 1980), to its maximum surface flux ({\it i.e.}, the flux of the brightest pixel). Specifically, in terms of SExtractor output quantities, for each detection we define the quantity $\mu$ as  \texttt{MU\_MAX $-$ MAG\_AUTO} minus the median value of \texttt{MU\_MAX $-$ MAG\_AUTO} for bright unsaturated stars on the image (note that the ratio of fluxes is equivalently the difference in magnitudes).

Unlike full-width at half maximum (FWHM), which is based on a one-dimensional symmetric ({\it e.g.}, Gaussian) PSF model, $\mu$ is versatile as a metric in that it does not involve any assumption of symmetry (radial or otherwise). Note that in Section 2 we defined and excluded radiation hit candidates as those detections having $\mu<-1$. A negative $\mu$ means the object is more concentrated than bright stars on the image, while a positive $\mu$ means it is more extended. The error in $\mu$, denoted $\sigma_\mu$, is obtained by adding in quadrature the instrumental magnitude error \texttt{MAGERR\_AUTO} and the 16th-to-84th percentile spread in \texttt{MU\_MAX $-$ MAG\_AUTO} for the bright stars.

\subsection{Systematic (non-cometary) variation in $\mu$}

The $\mu$ of a given small-body detection varies systematically with several known quantities, meaning that ``extended'' as defined by $\mu$ is not synonymous with ``cometary''.

Firstly, we must consider the apparent magnitude, since detections near the survey's limiting magnitude have a known bias \citep{ofe12a} in their instrumental magnitude (\texttt{MAG\_AUTO}), which by definition affects the value of $\mu$. \cite{ofe12a} note that use of the aperture magnitude \texttt{MAG\_APER} rather than the adaptive Kron magnitude \texttt{MAG\_AUTO} removes this bias, but unfortunately photometric zeropoints only exist presently for the latter in the PTF photometric database.

Secondly, the object's apparent motion on the sky during the 60-second exposure time must be considered, as such motion causes streaking to occur, which alters the flux distribution and hence also $\mu$. It turns out that in $>99$\% of all observations in our sample (mostly main-belt objects), the on-sky motion is smaller than $1''$/minute. Given PTF's $1''$/pixel resolution, one might expect that the vast majority of objects are not drastically affected by streaking. Nevertheless, $\mu$ varies systematically with motion, as it does for apparent magnitude (see Section 5.4). 

Thirdly, the photometric quality of an observation's host image, {\it i.e.}, the seeing (median FWHM) and sky brightness, must be taken into account, since the median and spread of  \texttt{MU\_MAX $-$ MAG\_AUTO} for bright stars on the image, and hence also $\mu$, are influenced by such conditions.

A final measurable property affecting $\mu$ is the distance between the object's flux barycenter and the center of its brightest pixel. In terms of SExtractor quantities, this is computed as
((\texttt{XPEAK\_IMAGE $-$ X\_IMAGE})$^2$ $+$ (\texttt{YPEAK\_IMAGE $-$ Y\_IMAGE})$^2$)$^{1/2}$. In particular, if the barycenter lies near to the pixel edge ($\sim$$0.5''$ from the pixel center), the majority of the flux will be nearly equally shared between two adjacent pixels. If it is near the pixel corner ($\sim$$0.7''$ from the center), the flux will be distributed into four pixels (assuming a reasonably symmetric PSF and non-Poisson-noise-dominated signal). An object's position relative to the pixel grid is random, but the resulting spread in barycenter position does cause systematic variation in $\mu$.

We can reasonably assume that some of these variations may be correlated. \cite{jed02} discusses systematic observable correlations of this kind, in the separate problem of debiasing sky-survey small-body data sets. Jedicke et al. also introduces a general formalism for representing survey detection systematics, which we now adapt in part to the specific problem of using $\mu$ to identify cometary activity.

\subsection{Formalism for interpreting $\mu$}

Let the state vector $\vec{x}$ contain all orbital ({\it e.g.}, semi-major axis, eccentricity) and physical ({\it e.g.}, diameter, albedo) information about an asteroid. Given this $\vec{x}$ there exists a vector of observed quantities $\vec{o}=\vec{o}(\vec{x})$. Most of these observed quantities are a function of the large number of parameters defining the sky survey (pointings, exposure time, optics, observatory site, data reduction, etc.). Included in $\vec{o}$ are the apparent magnitudes, on-sky motion, host-image seeing and sky brightness, barycenter-to-max-pixel distance, and also counts of how many total detections and how many unique nights the object is observed. In the above paragraphs we argued qualitatively that $\mu=\mu(\vec{o})$. 

Now suppose that $\vec{x}\to\vec{x}+\delta\vec{x}$, where addition of the perturbing vector $\delta\vec{x}$ is equivalent to the asteroid exhibiting a cometary feature. For instance, $\delta\vec{x}$ could contain information on a mass-loss rate or the physical (3-dimensional) scale of a coma or tail. The resulting change in the observables is

\begin{equation}
\vec{o}\to\vec{o}+\delta\vec{o}\quad\text{where}\quad\delta\vec{o}=\frac{d\vec{o}}{d\vec{x}}\cdot\delta\vec{x}
\end{equation}

\noindent The observation-perturbing vector $\delta\vec{o}$ could contribute to increased apparent magnitudes while leaving other observables such as sky position and apparent motion unchanged. To ``model'' the effect of cometary activity $\delta\vec{x}$ on the observables, {\it e.g.}, as in \cite{son11}, is equivalent to finding (or inverting) the Jacobian $d\vec{o}/d\vec{x}$, though this is unnecessary for the present analysis. The resulting change in the scalar quantity $\mu$ is

\begin{equation}
\mu\to\mu+\delta\mu\quad\text{where}\quad\delta\mu=\nabla\mu\cdot\delta\vec{o}
\end{equation}

Now suppose that some component of $\delta\vec{o}$ is (linearly) independent of $\vec{o}$, {\it i.e.}, there exists some unit vector $\hat{\imath}$ such that $\hat{\imath}\cdot\delta\vec{o}=\delta o_i>0$ while $\hat{\imath}\cdot\vec{o}=0$. Another way of stating this is that there exists some observable $o_i$ (the $i^{\rm{th}}$ component of $\vec{o}$), the value of which unambiguously discriminates whether the object is cometary or inert. An example would be the object's angular size on the sky. This need not be a known quantity; {\it e.g.}, in the case of angular size one would need to employ careful PSF deconvolution to accurately measure it. The details of $o_i$ do not matter, more important is its ability to affect $\mu$, as described below.

Given the existence of this discriminating observable $o_i$, we can write

\begin{equation}
\delta\mu=\nabla\mu\cdot\delta\vec{o}=\delta\mu_{\text{sys}}+\frac{\delta\mu}{\delta o_i}\delta o_i
\end{equation}

\noindent where the first term on the right side, $\delta\mu_{\rm{sys}}$ , represents systematic change in $\mu$ due to variation in known observables such as apparent magnitude and motion, and the second term represents a uniquely cometary contribution to $\mu$. We assume that $\delta\mu/\delta o_i\ne0$ in order for this reasoning to apply.

From our large sample of small-body observations, we are able to compare two objects, $\vec{o}$ and $\vec{o}\,'$, that have the same apparent magnitude, motion, seeing, etc. The computed $\delta\mu=\mu(\vec{o})-\mu(\vec{o}\,')$ in such a case must have $\delta\mu_{\rm{sys}}=0$, meaning a result of $\delta\mu\ne 0$ would imply one of the objects is cometary. We can then use prior knowledge, {\it e.g.}, that $\vec{o}$ is an inert object, to conclude $\vec{o}\,'$ is a cometary observation.

\subsection{A model-$\mu$ to describe inert objects}

\begin{figure*}
\centering
\includegraphics[scale=0.28]{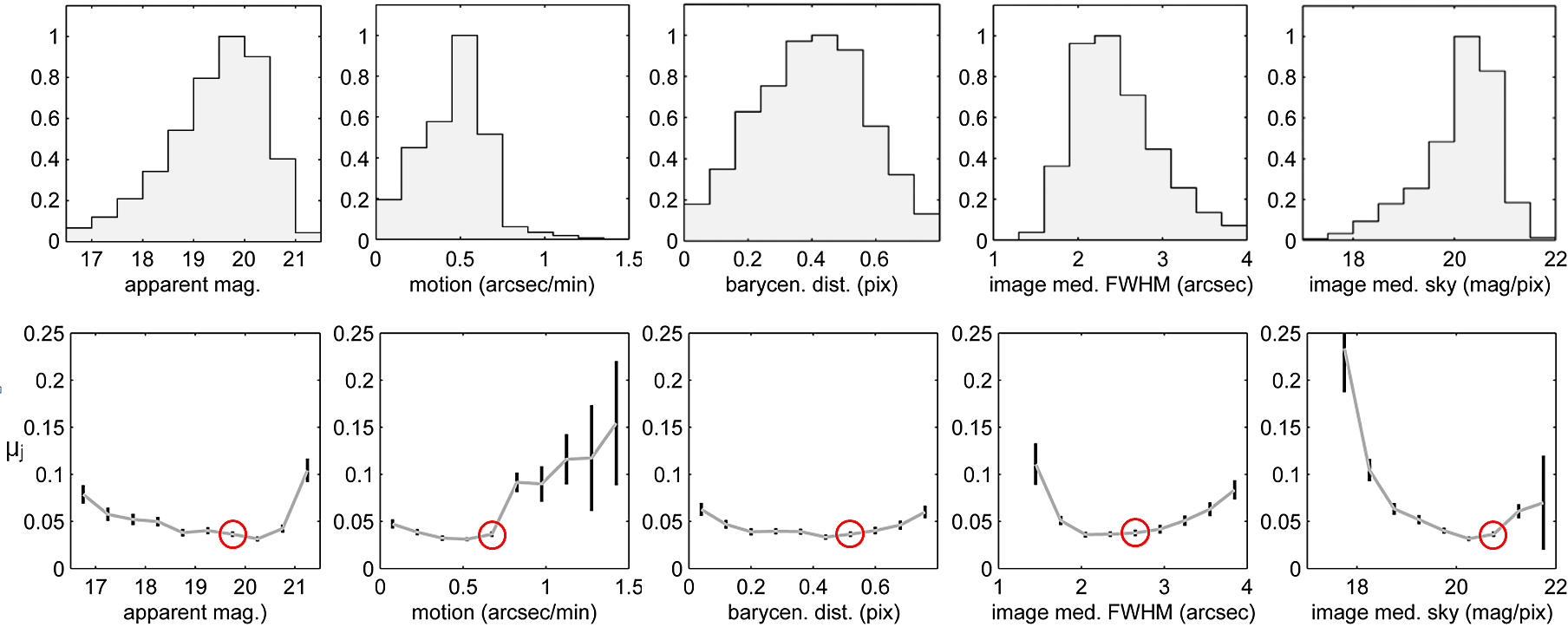}
\caption{Distribution of PTF asteroid observations in the five parameters comprising the observable vector $\vec{o}$. Bottom row: Model-$\mu$ values ($\mu_j$-values) plotted as a function of each parameter, while holding all other parameters constant; black bars show the error $\sigma_{\mu,j}$. The red circled point is the value at which the parameter is evaluated in the other plots. These plots only show a small slice of the much larger (and impossible to visualize) five-dimensional gridded function $\mu_j$.}
\end{figure*}

We build upon this formalism by employing prior knowledge of the apparent scarcity of main-belt comets. That is, we hypothesize that the vast majority of known objects in our sample are in fact inert, or mapped to an equivalently inert set of observations $\vec{o}$ when subjected to the survey mapping $\vec{x}\to\vec{o}(\vec{x})$. This allows us to construct a gridded model of $\mu$ for inert objects, denoted $\mu_j$.

We first bin the data in a five-dimensional $\vec{o}$-space and then compute the error-weighted mean of $\mu$ in each bin. The $j^{\rm{th}}$ bin in this $\vec{o}$-space is defined as the five-dimensional “box” having corners $\vec{o}_j$ and $\vec{o}_j+\Delta\vec{o}$. The model value $\mu_j$ in this $j^{\rm{th}}$ bin is found by summing over all observations in that bin:

\begin{equation}
\mu_j=\sigma_{\mu,j}^2\sum_{\vec{o}\in[\vec{o}_j,\vec{o}_j+\Delta\vec{o}]}\frac{\mu(\vec{o})}{\sigma_\mu(\vec{o})^2}
\end{equation}

\noindent where the scatter (variance) in the $j^{\rm{th}}$ bin is

\begin{equation}
\sigma_{\mu,j}^2=\left(\sum_{\vec{o}\in[\vec{o}_j,\vec{o}_j+\Delta\vec{o}]}\frac{1}{\sigma_\mu(\vec{o})^2}\right)^{-1}
\end{equation}

\noindent and the individual observation errors $\sigma_\mu$ are computed as described in Section 5.1. We exclude known comets from all bin computations, even though their effect on the mean would likely be negligible given their small population relative to that of asteroids. 

The histograms in Figure 9 show the range of values for the five components of $\vec{o}$, each of which is sampled in 10 bins. The five-dimensional $\vec{o}$-space considered thus has $10^5$ bins. However, given the centrally-concentrated distributions of each observable, only a fraction ($\sim$40\%) of these bins actually contain data points. Some bins ($\sim$9\%) only include a single data point; these data cannot be corrected using this $\mu_j$ model, but their content represent $<1$\% of the data. Lastly, $\sim$7\% of the data lie outside one or more of these observable ranges, and hence also cannot be tested using the model. Most of these excluded data are either low quality (seeing $> 4''$) or bright objects ($> 16.5$ mag).
Of the 175 previously known comets (see Section 3) plus 4 new (see Section 4) comets we found in PTF, 115 of these (64\%, mostly the dimmer ones) lie in these observable ranges and hence can be tested with the model.

\subsection{Defining a visually-screenable sample}

For each of the $\sim$2 million small-body observations in our data set, we use the inert model $\mu_j$ to define the corrected extendedness as a ``$\mu$-excess'':

\begin{equation}
\delta\mu=\mu-\mu_j
\end{equation}

\noindent and an uncertainty:

\begin{equation}
\sigma=\sqrt{\sigma_\mu^2+\sigma_{\mu,j}^2}
\end{equation}

\noindent For each of the $\sim$220,000 unique objects in our data set, we sum over all observations of that object to define

\begin{equation}
\overline{\delta\mu}=\langle\delta\mu\rangle^2\mathlarger{\sum}_{\begin{array}{cc}&\text{\footnotesize object's}\\[-1.25ex] &\text{\footnotesize observ-}\\[-1.25ex] &\text{\footnotesize ations}\end{array}}\frac{\delta\mu}{\sigma^2}
\end{equation}
\begin{equation}
\langle\delta\mu\rangle=\bigg(\mathlarger{\sum}_{\begin{array}{cc}&\text{\footnotesize object's}\\[-1.25ex] &\text{\footnotesize observ-}\\[-1.25ex] &\text{\footnotesize ations}\end{array}}\frac{1}{\sigma^2}\bigg)^{-1/2}
\end{equation}

These two quantities, $\overline{\delta\mu}$ and $\langle\delta\mu\rangle$, are useful for screening for objects which appear cometary in most observations. If an object is observed frequently while inactive but sparsely while active, $\overline{\delta\mu}$ and $\langle\delta\mu\rangle$ are less useful. As noted in Section 3.3, high cadence data are uncommon in our sample, alleviating this problem (see also Figure 15 in Section 7 for commentary on orbital coverage).

To select the sample to be screened by eye for cometary activity, we use the quantity $\overline{\delta\mu}-\langle\delta\mu\rangle$. In the case of normally-distributed data, the probability that this quantity is positive is $1-\rm{erf}(1)\approx 0.16$. As shown in Figure 10, the fraction of objects with $\overline{\delta\mu}-\langle\delta\mu\rangle>0$ is actually 0.007 (1,577 objects), much smaller than the Gaussian-predicted 0.16. This likely results from overestimated ({\it i.e.}, larger-than-Gaussian) $\langle\delta\mu\rangle$ values caused by outliers in the data. However, of the 115 testable comets in our data (111 known plus 4 new---see Table 5 for further explanation), 76 of these (66\%) have $\overline{\delta\mu}-\langle\delta\mu\rangle>0$. That is, a randomly chosen known comet from our sample is $\sim$100 times more likely to have $\overline{\delta\mu}-\langle\delta\mu\rangle>0$ than a randomly chosen asteroid from our sample, suggesting the criterion $\overline{\delta\mu}-\langle\delta\mu\rangle>0$ is a robust indicator of cometary activity. 

The fact that only 66\% of the 115 comets in our testable sample satisfy $\overline{\delta\mu}-\langle\delta\mu\rangle>0$ means that, if one assumes main-belt comets share the same extendedness distribution as all comets, then our detection method is only 66\% efficient. Sufficiently weak and or unresolved (very distant) activity inevitably causes the lower and negative values of $\overline{\delta\mu}-\langle\delta\mu\rangle$.

\begin{figure}
\centering
\includegraphics[scale=0.11]{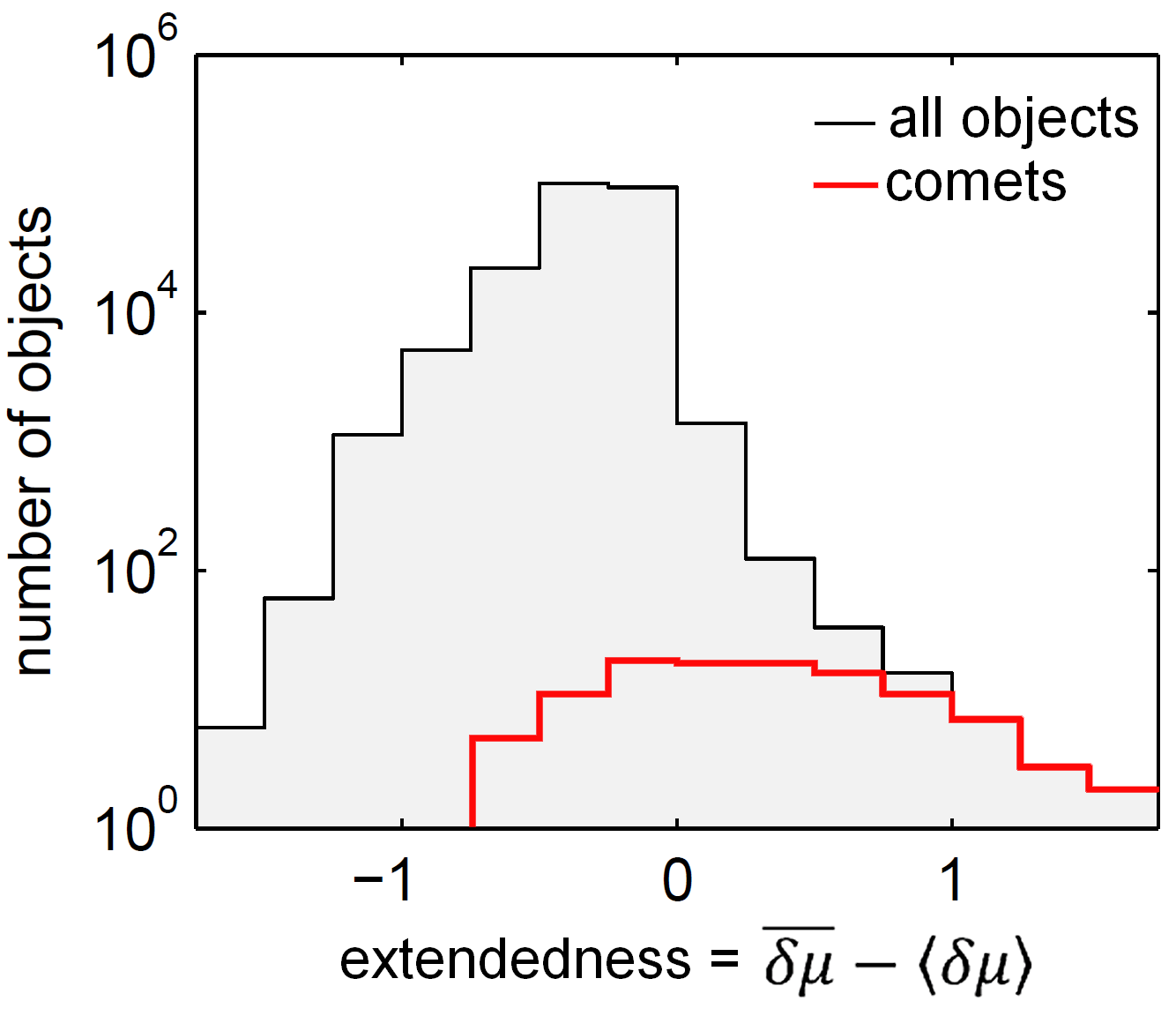}
\caption{For inclusion in the screening sample, an object's mean extendedness value, $\overline{\delta\mu}$, must exceed zero by more than one-sigma, $\langle\delta\mu\rangle$, which in this histogram is true for all objects to the right of zero. The 111 known (and testable) comets plus the 4 new comets (see Section 4.3) are plotted in red; 76 out of these 115 fall to the right of zero, meaning our method is 76/115 = 66\% efficient at detecting the known plus new comets comprising our sample. Both of the MBCs in our sample fall to the right of zero, implying that we are 100\% efficient at detecting objects at least as extended as these known MBCs.}
\end{figure}

Given the specific goal to detect main-belt objects that are at least as active as the known candidate MBCs, we consider the value of $\overline{\delta\mu}-\langle\delta\mu\rangle$  for the known candidate MBCs in our sample. P/2010 R2 (La Sagra) has $\overline{\delta\mu}-\langle\delta\mu\rangle=0.474$ (from 34 observations made on 21 nights). P/2006 VW$_{139}$ has $\overline{\delta\mu}-\langle\delta\mu\rangle=0.231$ (from 5 observations made on 3 nights). Hence, the $\overline{\delta\mu}-\langle\delta\mu\rangle>0$ criterion is more than sufficient (formally, 100\% efficient) for detecting extendedness at the level of these known, kilometer-scale candidate MBCs. Note however that we do \emph{not} claim 100\% detection efficiency with respect to objects of similar \emph{magnitude} as these candidate MBCs; see Figure 11 for a consideration of efficiency as a function of apparent magnitude. 

\section{Extended-object analysis: Results}

A total of 1,949  observations (those having $\delta\mu-\sigma>0$) of 1,577 known and newly discovered objects satisfying $\overline{\delta\mu}-\langle\delta\mu\rangle>0$ were inspected visually to identify either contamination from image artifacts or true cometary features. For each detection this involved viewing a $2'\times2'$ cutout of the image, with contrast stretched from $-0.5\sigma$ to $+7\sigma$ relative to the median pixel value (where $\sigma=\sqrt{\rm{median}}$). This image was also flashed with the best available image of the same field taken on a different night (``best'' meaning dimmest limiting magnitude), to allow for rapid contaminant identification.

With the exception of two objects (described below), virtually all of these observations were clearly contaminated by either a faint or extended nearby background source, CCD artifacts or optical artifacts (including ghosts and smearing effects). In principle these observations should have been removed from the list of transients by the filtering process described in Section 2.2, however some residual contamination was inevitable. 

The screening process did however reveal cometary activity on two non-main-belt objects previously labeled as asteroids: 2010 KG$_{43}$ and 2011 CR$_{42}$, which had $\overline{\delta\mu}-\langle\delta\mu\rangle$ values of 0.2 and 1.1, respectively (Figure 12). Note that taking these two objects into account improves our efficiency slightly to $(76+2)/(115+2)=67$\%.

In addition to the two known candidate MBCs (Figure 13)---which were among the 76 comets already noted to have passed the screening procedure---this process also confirmed the extendedness of three of the four comets discovered in PTF as moving-objects (2009 KF$_{37}$, 2010 LN$_{135}$, and 2012 KA$_{51}$) as described in Section 4. These comets had $\overline{\delta\mu}-\langle\delta\mu\rangle$ values of 0.33, 0.29, and 0.58, respectively. The procedure did \emph{not} identify  the fourth new comet discovery, C/2012 LP$_{26}$ (Palomar), as an extended object, suggesting that it was unresolved.

\begin{figure}
\centering
\includegraphics[scale=0.15]{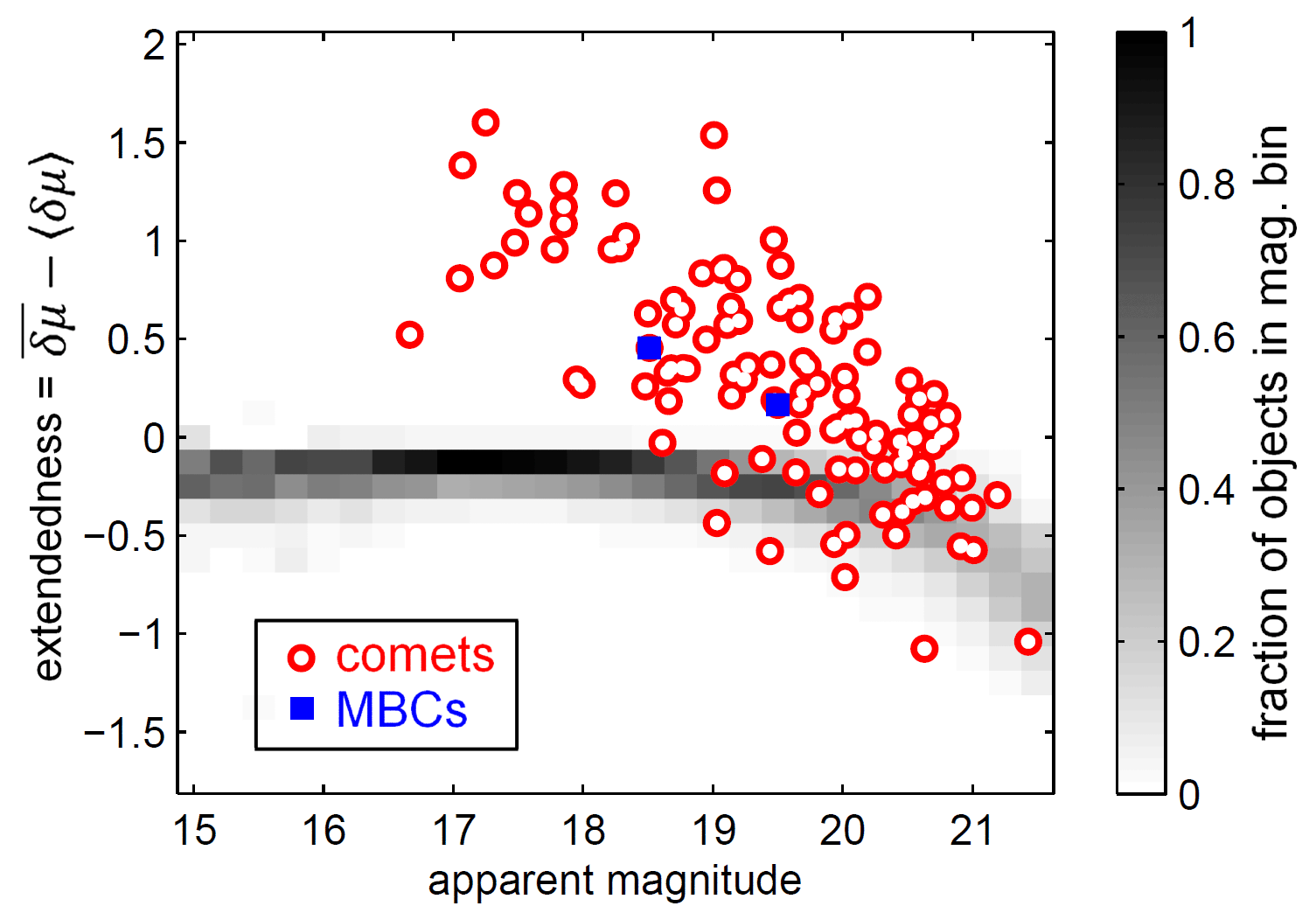}
\caption{Like Figure 10, but now with each object's median apparent magnitude plotted as well. All small bodies in the sample are represented in the 2D histogram (normalized with respect to each magnitude bin), while the known comets are overplotted as red-white circles, the two MBCs as blue squares. Again, about two-thirds (66\%) of the comets lie above zero. This plot suggests that the completeness is expressible as a function of apparent magnitude. That is, $C>66\%$ for bright comets and $C<66\%$ for dim comets (approaching zero for $>21$ mag), while on average $C=66\%$. The exact magnitude dependence is sensitive to bin size and is not explored quantitatively here.}
\end{figure}

\begin{table*}
 \renewcommand\thetable{4}
\caption{Comets discoveries made in PTF in the course of this work}
\hfill{}
\begin{tabular}{lccccllcr}
\hline\\[-3ex]
name&$q$ (AU)&$e$&$i$ (deg)&$T_{\rm{peri}}$&dates in PTF&type$^*$&co-discoverer&Reference$^{\dagger}$\\
\hline
2009 KF$_{37}$&2.59&0.34&11.2&2009-Aug&2009-May to 2009-Jul&JF&---&MPS 434214\\
2010 KG$_{43}$&2.89&0.49&13.5&2010-Jul&2010-Aug to 2010-Sep&JF&WISE (discovered orbit)&MPS 434201\\
2010 LN$_{135}$&1.74&1.00&64.3&2011-May&2010-Jun to 2010-Jul&LP&---&MPS 439624\\
2011 CR$_{42}$&2.53&0.28&8.5&2011-Nov&2011-Mar&JF&Catalina (discovered orbit)&CBET 2823\\
2012 KA$_{51}$&4.95&1.00&70.6&2011-Nov&2012-May&LP&---&MPS 434214\\
C/2012 LP$_{26}$ (Palomar)&6.53&1.00&25.4&2015-Aug&2012-Jun to 2012-Jul&LP&Spacewatch (discovered coma)&CBET 3408\\
\hline
\end{tabular}
\smallskip
\\$^*$JF = Jupiter-family; LP = long-period\;\;\;$^{\dagger}$MPS = {\it Minor Planet Circulars Supplement}; CBET = {\it Central Bureau Electronic Telegram}
\hfill{}
\end{table*}

\subsection{A new quasi-Hilda comet: 2011 CR$_{42}$}

The object 2011 CR$_{42}$, discovered on 2011-Feb-10 by the Catalina Sky Survey \citep{dra09}, has an uncommon orbit ($a = 3.51$ AU, $e = 0.28$ and $i = 8.46^{\circ}$). Six PTF $g'$-band observations made between 2011-Mar-05 and Mar-06 \citep{was11} all show a coma-like appearance but no tail. The object was 2.92 AU from the Sun and approaching perihelion ($q = 2.53$ AU on 2011-Nov-29). Based on its orbit and using IAU phase-function parameters \citep{bow89} $H = 13.0$ and $G = 0.15$, 2011 CR$_{42}$ should have been easily observed at heliocentric distances 3.8 AU and 3.1 AU in 2010-Feb and 2010-Dec PTF data, with predicted magnitudes of 19.2 and 18.8 mag, respectively. Upon inspection of these earlier images, no object was found within $200''$ of the predicted position. Its absence in these images further suggests cometary activity.

Like the MBCs and unlike most Jupiter-family comets, 2011 CR$_{42}$'s Tisserand parameter \citep{mur99} with respect to Jupiter ($T_{\rm{Jup}} = 3.042$) is greater than 3. While the criterion $T_{\rm{Jup}}>3$ is often used to discriminate MBCs from other comets, we note that MBCs more precisely have $T_{\rm{Jup}}>3.1$. About half of the $\sim$20 \emph{quasi-Hilda comets} (QHCs, \citealp{tot06} and refs. therein) have  $3<T_{\rm{Jup}}<3.1$, as does 2011 CR$_{42}$. Three-body (Sun + Jupiter) interactions tend to keep $T_{\rm{Jup}}$ approximately constant (this is akin to energy conservation). Such interactions nonetheless can chaotically evolve the orbits of QHCs. Their orbits may settle in the stable 3:2 mean-motion (Hilda) resonance with Jupiter at 4 AU, wander to a high-eccentricity Encke-type orbit, or scatter out to (or in from) the outer Solar System. Main-belt orbits, however, are inaccessible to these comets under these $T_{\rm{Jup}}$-conserving three-body interactions.

To verify this behavior, we used the hybrid symplectic integrator MERCURY \citep{cha99} to evolve 2011 CR$_{42}$'s orbit forward and backward in time to an extent of $10^4$ years. For initial conditions we tested all combinations of 2011 CR$_{42}$'s known orbital elements plus or minus the reported error in each (a total of $3^6 = 729$ runs in each direction of time). We did not include non-gravitational (cometary) forces in these integrations, as it was assumed that this object's relatively large perihelion distance would render these forces negligible. In $\sim$25\% of the runs, the object scattered out to (or in from) the outer solar System in less than the $10^4$ year duration of the run. In the remainder of the runs, its orbit tended to osculate about the stable 3:2 mean motion Jupiter resonance at 4 AU. These results strongly suggest that 2011 CR$_{42}$ is associated with the Hilda family of objects belonging to this resonance, and thus likely is a quasi-Hilda comet.

\begin{figure}
\centering
\includegraphics[scale=0.16]{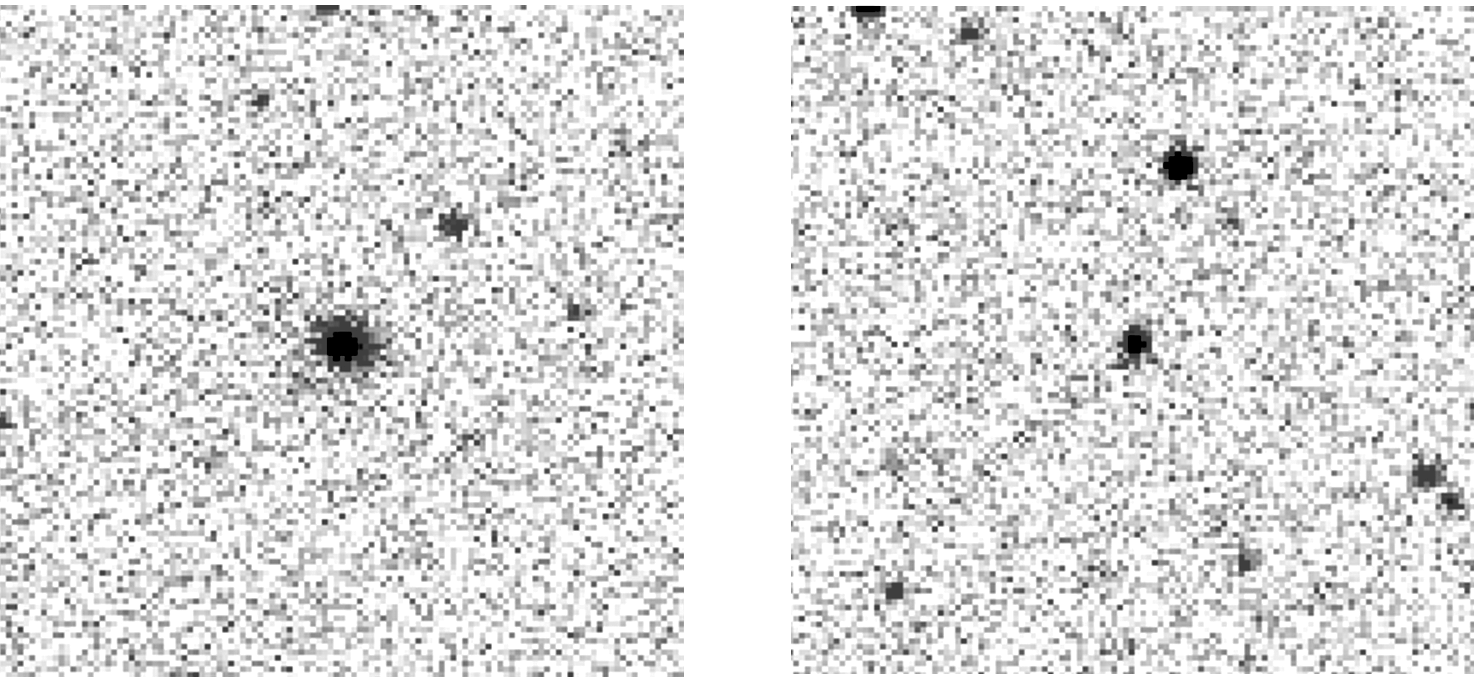}
\caption{Known asteroidally-designated objects whose cometary activity was discovered in PTF in the course of this work. Each image is $2'\times2'$ (pixel scale $1.01''$).  Left: 2011 CR$_{42}$ in $g'$-band on 2011-03-06. No tail is discernible, but the object's FWHM is twice that of nearby stars. Right: 2010 KG$_{43}$ in $R$-band on 2010-09-08. A $\sim$$1'$-long tail is discernible extending toward the lower left corner of the image. }
\end{figure}

\begin{figure}
\centering
\includegraphics[scale=0.16]{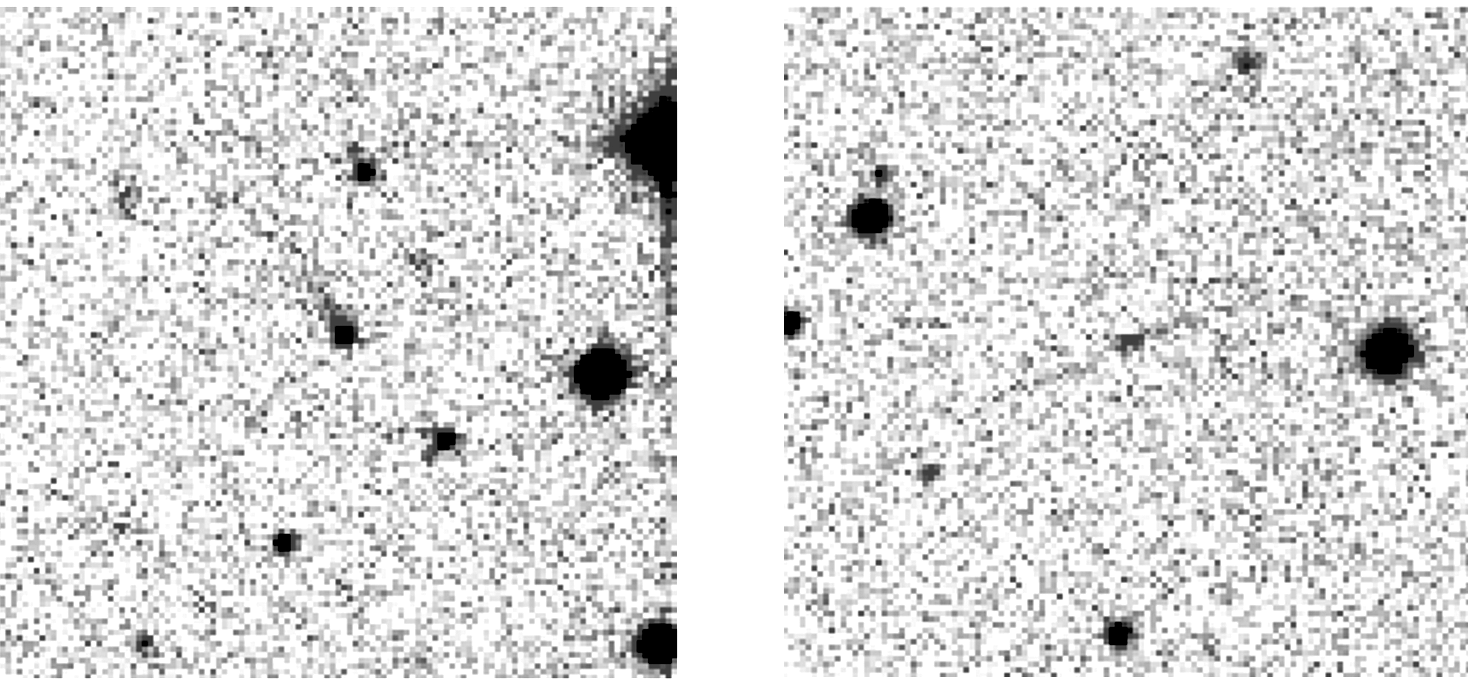}
\caption{Known candidate main-belt comets in PTF. Each image is $2'\times2'$ (pixel scale $1.01''$).  Left: P/2010 R2 (La Sagra) in $R$-band ($R\sim18.5$ mag) on 2010-08-19, with its tail extending towards the top left. Right: P/2006 VW$_{139}$ in $g'$-band ($g'\sim20$ mag) on 2011-12-21, with its two oppositely oriented tails barely discernible by eye.}
\end{figure}

\begin{table}

 \renewcommand\thetable{5}
\centering
\renewcommand{\arraystretch}{1.2}
\caption{Summary of the PTF comet sample. JF = Jupiter-family; LP = long-period; MB = main-belt. ``Observed'' means found by the search algorithms of Sections 3 or 4; ``model-$\mu$ tested'' means it lies in the observable ranges shown in Figure 9 ({\it e.g.}, excludes bright comets), and  $\overline{\delta\mu}-\langle\delta\mu\rangle>0$ means positively detected as extended. Objects 2010 KG$_{43}$ and 2011 CR$_{42}$ are counted in all rows as PTF-discovered JFCs, even though they were not included in the $76/115=66$\% efficiency calculation of Section 5.5 (since they were not discovered until Section 6). }
\hfill{}
\begin{tabular}{rccccccc}
\hline
&\multicolumn{3}{c}{previously known}&\multicolumn{3}{c}{PTF discovered}&\\\cline{2-7}
&JF&LP&MB&JF&LP&MB&total\\[-0.5ex]
\hline
observed&108&65&2&3&3&0&181\\[-0.5ex]
model-$\mu$ tested&71&38&2&3&3&0&117\\[-0.5ex]
$\overline{\delta\mu}-\langle\delta\mu\rangle>0$&44&27&2&3&2&0&78\\[-0.5ex]
\hline
\end{tabular}
\hfill{}
\end{table}

\section{Statistical interpretation}

\subsection{Bayesian formalism}

Following the approach of \cite{son11} and borrowing some of their notation, we apply a Bayesian formalism to our survey results to estimate an upper limit for the fraction $f$ of objects  (having $D>1$ km) which are active MBCs at the time of observation. The prior probability distribution on $f$ is chosen to be a log-constant function:

\begin{equation}
P(f) = -\frac{1}{f\text{log}f_{\text{min}}} \;\;\text{for}\;\;f_{\text{min}}<f<1 \\ 
\end{equation}

\noindent This prior is justified since we know $f$ is ``small'', but not to order-of-magnitude precision. The minimum value $f_{\text{min}}>0$ is allowed to be arbitrarily small, since the integral of $P(f)$ is always unity:

\begin{equation}
\int_{f_{\text{min}}}^1 P(f) \; df = 1\\ 
\end{equation}

\noindent Let $ N$ be the number of objects in a given sample, $n$ the number of active MBCs positively detected in that sample, and $C$ the completeness or efficiency of our MBC-detection scheme. In Section 5.5 we discussed how $C=0.66$ if defining completeness with respect to the extendedness distribution of the 115 known comets on which we tested our detection method. Relative to objects at least as extended as the two known candidate MBCs we tested, however, we can take $C=1$ (100\% efficiency), since both of the MBCs observed were robustly flagged by our screening process.

The likelihood probability distribution function for a general sample $S$ is formally a binomial distribution, but because the samples we will be considering are very large ($N\gg1$), the likelihood function is well-approximated as a Poisson distribution:

\begin{equation}
\begin{aligned}
\setlength{\jot}{14pt} 
P(S|f)&=\frac{N!}{n!(N-n)!}(Cf)^n(1-Cf)^{N-n}\\
&\approx\frac{(NCf)^n}{n!}\exp(-NCf)
\end{aligned}
\end{equation}

\noindent  Bayes' Theorem then gives the formula for the posterior probability distribution on $f$ given our results:

\begin{equation}
P(f|S)=\frac{P(S|f)P(f)}{\int_{f_{\text{min}}}^1 P(S|f)P(f)\;df}\propto f^{n-1}\exp(-NCf)
\end{equation}

\noindent The constant of proportionality (not shown) involves incomplete gamma functions\footnote{The \emph{incomplete gamma function} is defined as \[ \Gamma(n,N)=\int_N^\infty t^{n-1}\exp(-t)\;dt\]}, and is well-defined and finite (including in the limit $f_{\text{min}}\to0$).

Finally, we can compute the 95\% confidence upper limit $f_{95}$ by solving the implicit equation

\begin{equation}
\int_0^{f_{95}}P(f|S)\;df=0.95
\end{equation}

\subsection{Active MBCs in the \emph{entire} main-belt}

\noindent We first take the sample $S$ to be representative  of \emph{all} main-belt asteroids, which in our survey amounted to $N=2.2\times10^5$ observed objects and $n=2$ detected MBCs. Equation (14) yields 95\% confidence upper limits for $f$ of $33\times10^{-6}$ and $22\times10^{-6}$, for efficiency values of $C$ of 0.66 and 1.0, respectively. Figure 14 depicts the probability distributions for each case.

We note that although these results are based on the positive detection of only two candidate MBCs, the reader need not be skeptical on the basis of ``small number statistics'', since the Possionian posterior (Equation 13 and Figure 14) formally accounts for ``small number statistics'' through its functional dependence on $n$. Even if we had detected no MBCs at all---in which case $n$ would be zero (as was the case in \citealp{son11})---the posterior would still be well-defined; the 95\%-confident upper limit would naturally be larger to reflect the greater uncertainty.

Our discussion has so far only considered the \emph{fraction} of active MBCs, rather than their \emph{total number}. This is because an estimate of the total underlying number of main-belt objects (down to $D\sim1$ km) must first be quoted from a properly-debiased survey. A widely-cited example is \cite{jed98}, who applied a debiasing analysis to the Spacewatch survey and concluded that there are of order $10^6$ main-belt asteroids (down to $D\sim1$ km)\footnote{More recent survey results will eventually test/verify this result, {\it e.g.} the WISE sample has already produced a raw size-frequency distribution \citep{mas11}, the debiased form of which will be of great value.}. Since the MBC-fraction estimates in the above paragraphs are conveniently given in units of per million main-belt asteroids, we directly estimate the upper limit on the total number of active MBCs in the true underlying $D>1$ km population (again to 95\% confidence) to be between 33 and 22, depending on the efficiency factor $C$ (0.66 or 1.0).

\begin{figure}
\centering
\includegraphics[scale=0.16]{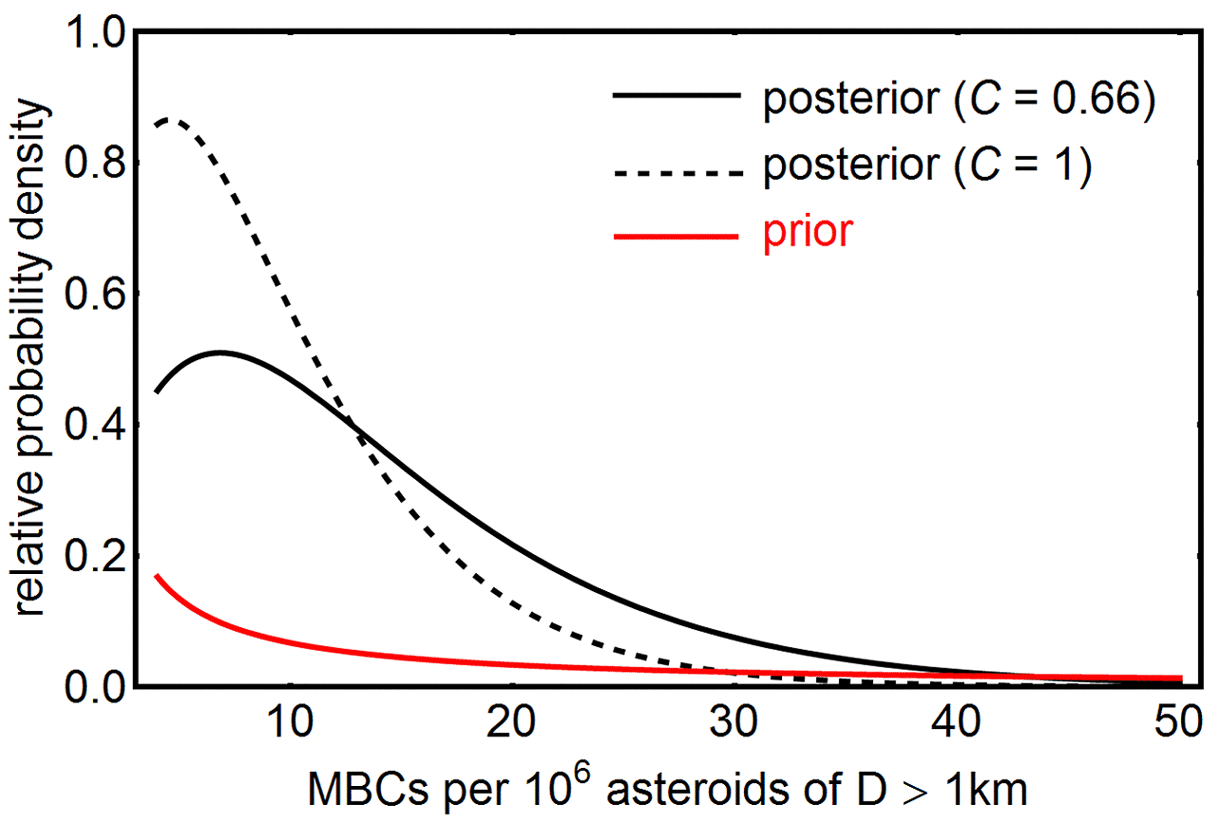}
\caption{Probability distributions for estimating the fraction of main-belt comets, based on the results of our sample screening. The $C=0.66$ case reflects our efficiency with respect to the extendedness distribution of all known comets, while the $C=1$ case applies to extendedness levels at least as high as the two known candidate MBCs in our sample. In this plot $f_{\text{min}}$ was set to $4\times10^{-6}$, to facilitate visual comparison of the normalized prior with the normalized posteriors.}
\end{figure}

\subsection{Active MBCs in the \emph{outer} main-belt}

Of the seven candidate MBCs listed in Table 1, all except 259P/Garradd have semi-major axes between 3.0 AU and 3.3 AU, corresponding approximately to the 9:4 and 2:1 Jupiter resonances (Kirkwood gaps). This semi-major axis constraint is satisfied by 123,366 ($\sim$20\%) of the known objects as of August 2012, of which 47,450 (38\%) are included in the PTF sample.

Reapplying equations (10)--(14) except now using $N=47,450$ (while $n=2$ remains unchanged), we find 95\%-confidence upper limits of 160 and 110 active MBCs per million outer main-belt asteroids with $3.0 < (a/\text{AU}) < 3.3$ and $D>1$ km, for detection efficiencies of $C=0.66$ and $C=1.0$, respectively. 

Although only $\sim$20\% of the \emph{known} main-belt asteroids lie in this orbital range, the debiased semi-major axis distribution presented in \cite{jed98} predicts that $\sim$30\% of all main-belt objects (of $D>1$ km) lie in this outer region. The discrepancy is due to the fact that these objects are more difficult to detect, since they are further away and tend to have lower albedos (this lower detection efficiency is evident for instance in the WISE sample shown in Figure 3). Assuming 300,000 objects actually comprise this debiased outer main-belt region, the inferred 95\% confidence upper limit on the total number of active MBCs existing in this region is $\sim$50 (for $C=0.66$) and $\sim$30 (for $C=1.0$).

\subsection{Active MBCs in the \emph{low-inclination} outer main-belt}

Four out of the seven candidate MBCs in Table 1 have orbital inclinations of $i<5^{\circ}$. Combined with the semi-major-axis constraint $3.0 < (a/\text{AU}) < 3.3$, this associates them with (or close to) the Themis asteroid family. There are 25,069 objects in the small-body list we used which satisfy this combined constraint on $a$ and $i$ ($\sim$4\% of the known main-belt), of which 8,451 (34\%) are included in the PTF sample.

Again reapplying equations (10)--(14), we now use $N=8,451$ and $n=1$, where the new value for $n$ reflects the fact that P/2006 VW$_{139}$ satisfies this $i$-criterion while P/2010 R2 (La Sagra) does not. We find 95\%-confidence upper limits of 540 and 360 active MBCs per million low-inclination, outer main-belt asteroids, for detection efficiencies of $C=0.66$ and $C=1.0$, respectively. 

Once again, \cite{jed98} offer estimates of the debiased number of objects in the underlying population of interest: for outer main-belt asteroids, they found that $\sim$20\% of the debiased objects had $i<5^{\circ}$. Hence, assuming there are 60,000 objects (of $D>1$ km) in the actual low-$i$ outer main-belt population satisfying these $a$ and $i$ constraints, the resulting upper limit estimates for the total number of active MBCs it contains is $\sim$30 (for $C=0.66$) and $\sim$20 (for $C=1.0$).

\subsection{Active MBCs among low-$i$ outer main-belt objects observed \emph{near perihelion} ($-45^\circ<\nu<45^\circ$)}

Of the 8,451 low-inclination outer main-belt objects observed by PTF (see Section 7.4), 5,202 were observed in the orbital quadrant centered on perihelion (in terms of true orbital anomaly $\nu$, this quadrant is $-45^\circ<\nu<45^\circ$). We consider this constraint given that all known candidate MBCs (Table 1) have shown activity near perihelion. Now using $N=5,202$ and $n=1$ (here again $n=1$ represents P/2006 VW$_{139}$), we find 95\%-confidence upper limits of 880 and 570 active MBCs per million low-inclination, outer main-belt asteroids observed by PTF near perihelion, for detection efficiencies of $C=0.66$ and $C=1.0$, respectively.

We caution that, unlike the previous subsets (which were defined solely by orbital elements), the population to which these statistics apply is less well-defined. In particular, the bias for detection near perihelion (Figure 15), due in part to the $(r\Delta)^{-2}$ dependence in the reflected sunlight, is more pronounced for smaller, lower-albedo, higher eccentricity objects. Hence, naively imposing a constraint on true anomaly $\nu$ implictly introduces selection biases in $D$, $p_V$ and $e$. Moreover, these implicit biases depend on the sensitivity of the PTF survey in a more nuanced manner, invalidating the simple $D\gtrsim1$-km lower limit we have quoted generally in this work. Nonetheless, these parameters ($D$ and $p_V$) are important enough to merit individual treatment, as detailed below.

\subsection{Active MBCs in the sub-5 km diameter population}

Yet another constraint that well-encompasses the known MBC candidates of Table 1 is a diameter $D<$ 5 km (corresponding to approximately $H>15$ mag for albedo $p_V=0.07$). Applying this constraint decreases the number of PTF-sampled objects $N$ by 28\%, 45\% and 41\% for the entire main-belt, outer main-belt, and low-$i$ outer main-belt, respectively. These smaller sample sizes result in slightly higher 95\%-confidence upper limits for the fraction of active MBCs: 30--45, 180--280, and 610--920 per $10^6$ objects having $5>(D/\text{km})>1$ in the entire main-belt, outer main-belt, and low-$i$ outer main-belt, respectively (the ranges corresponding to the two values of the efficiency factor $C$).

\cite{jed98} found that the debiased differential number distribution as a function of absolute magnitude $H$ is $\sim$$10^{\alpha H}$, where $\alpha\approx0.35$. The resulting cumulative number distribution ({\it i.e.}, the number of asteroids brighter than absolute magnitude $H$) is $\sim$$10^{\alpha H}/\alpha \ln(10)$. Using $H=17$ in this expression gives the predicted $10^6$ asteroids having $D>1$ km. The fraction of these objects in the range $15<H<17$ is therefore $1-10^{\alpha(15-17)}\approx80$\%. Scaling the debiased populations discussed above by this factor and using the new limits from the preceding paragraph gives new upper limits on the \emph{total} number of active MBCs existing in the three regions: $\sim$24--36, $\sim$40--70, and $\sim$30--45 in the entire main-belt, outer main-belt, and low-$i$ outer main-belt, respectively.

\subsection{Active MBCs among low-albedo (WISE-sampled) objects}

 \cite{bau12} analyzed WISE observations of five of the active-main-belt objects listed in Table 1. By fitting thermal models to the observations, they found that all of these objects had visible albedos of $p_V<0.1$. As shown in Figure 3 and described in Section 3.4, about half of the asteroids which were observed by WISE also appear in the PTF sample; in particular there were $N=32,452$ low-albedo ($p_V<0.1$) objects observed by both surveys. Included in the \cite{bau12} sample was PTF-observed candidate MBC P/2010 R2 (La Sagra), whose fitted albedo of $p_V=0.01\pm 0.01$ implies we can take $n=1$ (one positive active MBC-detection) in the low-albedo WISE/PTF sample.

Following the 95\% confidence upper limit computation method of the previous sections, we derive upper limits of 90--140 active MBCs per $10^6$ low-albedo ($p_V<0.1$) asteroids. As mentioned earlier, the full-debiasing of the WISE albedo distribution \citep{mas11} will eventually allow us to convert this upper limit on the \emph{fraction} of active MBCs among low-albedo asteroids into an upper limit on their \emph{total number}, just as \cite{jed98} has allowed us to do for orbital and size distributions.

\subsection{Active MBCs among C-type (SDSS-sampled) objects}

The MBC candidate P/2006 VW$_{139}$ was observed serendipitously by SDSS on two nights in September 2000. While one of the nights was not photometric in $g$-band, the other night provided reliable $g,r,i$ multi-color data on this object, yielding a principal component color $a^*=-0.14\pm0.05$. Because it has $a^*<0$, this suggests P/2006 VW$_{139}$ is a carbonaceous (C-type) object\footnote{This taxonomic classification for P/2006 VW$_{139}$ has been confirmed spectroscopically by \cite{hsi12b} and \cite{lic13}.}.

Figure 3 depicts the overlap of the SDSS-observed sample with PTF, which includes $N=24,631$ C-type ($a^*<0$) objects. Taking $n=1$, we derive 95\% confident upper-limits of 120--190 active MBCs per $10^6$ C-type asteroids (where again the range corresponds to $C=0.66$--$1.0$).

\section{Conclusion}

\subsection{Summary}

\begin{figure*}
\centering
\vspace{5pt}
\includegraphics[scale=0.17]{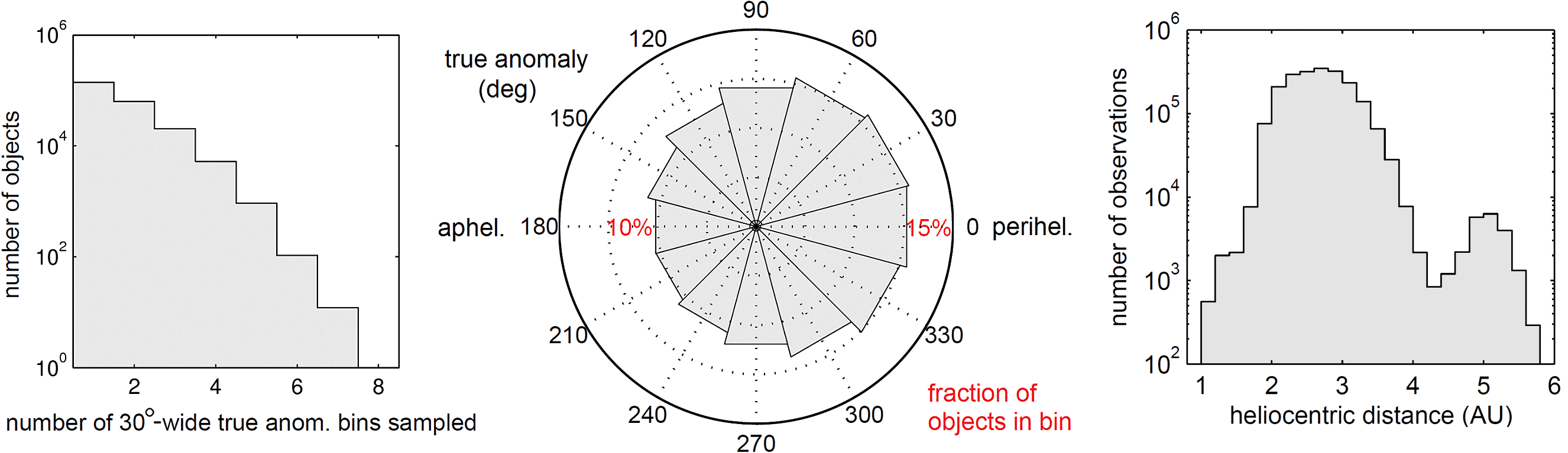}
\caption{Summary of true-anomaly and heliocentric distance coverage of known small bodies in PTF (see Figure 3 for other orbital statistics of this sample), from March 2009 through July 2012. Most ($\sim$90\%) of objects have only been sampled in at most two 30$^{\circ}$-wide true anomaly bins, {\it i.e.}, less than 1/6 of the orbit. In the left histogram objects are only counted once, but in the rose diagram (middle), each object is counted once for each bin it in which it is sampled (hence the fraction values reported for all twelve bins sum to more than 100\%). Although all objects spend more time around aphelion, most only fall above the survey detection limit near perihelion, hence there is a slightly larger fraction of objects observed near perihelion.}
\end{figure*}

Using original kd-tree-based software and stringent quality filters, we have harvested observations of $\sim$40\% ($\sim$220,000) of the known solar system small bodies and 626 new objects (622 asteroids and 4 comets) from the first 41 months of PTF survey data (March 2009 through July 2012). This sample is untargeted with respect to the orbital elements of known small bodies (but not necessarily the \emph{true} underlying population), down to $\sim$1-km diameter-sized objects. Most ($\sim$90\%) of the objects are observed on less than $\sim$10 distinct nights, and $\sim$90\% are observed over less than 1/6 of their orbit, allowing us to characterize this sample predominantly as a ``snapshot'' of objects in select regions of their orbits.

We have introduced a metric for quantifying the extendedness of a small body in an image, and have corrected this metric, on a per-observation basis, for systematic variation due to observables such as apparent magnitude, on-sky motion and pixel-grid alignment. In this metric, an extendedness of zero describes stellar-like (asteroidal) objects, whereas a positive value indicates potentially cometary extendedness.

We defined a sample for visual screening consisting of all objects whose mean extendedness value is greater than zero by at least one-sigma. This screening sample consisted of $\sim$1,500 unique objects, 76 (out of 115) comets, and two known candidate active MBCs, P/2010 R2 (La Sagra) and P/2006 VW$_{139}$, which upon inspection appear active and visibly extended in the images. Of the $\sim$1,500 objects screened, we found evidence for activity on two known (non-main-belt) asteroidally-designated objects, 2010 KG$_{43}$ and 2011 CR$_{42}$, and confirmed activity on the three out of the four (non-main-belt) comets that our moving-object algorithm discovered.

Given these results, using a log-constant prior we infer with 95\% confidence an upper limit of $<33$ active MBCs per $10^6$ main-belt asteroids for a $C=0.66$ detection efficiency with respect to the extendedness distribution of known comets, and $<22$ active MBCs per $10^6$ main-belt asteroids for a 100\% efficiency with respect to objects at least as extended as the known candidate active MBCs in our sample. 

\subsection{Comparison to previous work}

Our inferred 95\% confidence upper limit of at most $\sim$30 active MBCs per $10^6$ main-belt asteroids of $D>1$ km is comparable but slightly lower than that of Gilbert and Wiegert (\citeyear{gil09}, \citeyear{gil10}), who estimated $40\pm18$ active MBCs per $10^6$ main-belt asteroids, also for $D>1$ km, from visual inspection of a similarly untargeted sample of $\sim$25,000 objects from the \emph{Canada-France-Hawaii Telescope Legacy Survey} (Gilbert and Wiegert (\citeyear{gil09}, \citeyear{gil10}). That result was based on the detection of a single unknown comet in their sample, which was never actually confirmed to be a main-belt object due to lack of follow-up observations. Even before taking into account our order-of-magnitude larger sample size, we note that, in contrast to their results, our limits are based on positive MBC detections and use detection efficiencies estimated from observations of $\sim$100 known comets.

The result of \cite{son11} was a much larger upper limit of $\sim$3,000 MBCs per $10^6$ main-belt asteroids (to 90\% confidence), albeit applicable to the smaller limiting diameter of $\sim$0.5 km. Their smaller sample size of 924 objects is certainly the cause for their much larger uncertainty. While their detection methods were proven robust with respect to known candidate MBCs, we note that their sample included no unambiguously cometary objects. Hence, it is difficult to compare our result to theirs, but the possibility of a steeply increasing number distribution for MBCs below the $\sim$1-km level is not ruled out. Indeed, two known candidate MBCs, 238P/Read and 259P/Garradd, have measured sub-kilometer diameters (\citealp{hsi09b};  \citep{mac12}). 

\subsection{Future work: Photometric (absolute magnitude) variation as a function of orbital anomaly}

As suggested by this article's title, extended-object analysis is only the first kind of cometary-detection method to which we intend on subjecting the PTF small-body data set. In a planned Part II to this study, we hope to analyze the time- (and mean-anomaly-) varying absolute magnitude of small bodies over orbital-period baselines, which could potentially reveal even unresolved cometary activity.

Preliminary analysis of PTF  photometry of main-belt comet P/2010 R2 (La Sagra), which include pre-discovery observations, shows a time-resolved $\sim$1.5-mag increase in absolute magnitude and a corresponding factor $\sim$5 increase in the dust-to-nucleus cross-section ratio, $A_d/A_N$. These results suggest PTF is capable of detecting intrinsic disk-integrated flux variation at the level of known candidate MBCs. Upcoming analyses of other known comets in our sample should confirm this robustness.

As shown in Figure 15, the orbital-coverage of PTF-observed known objects is far from complete. The orbital period of main-belt objects varies from about three to six years; a desirable prerequisite to orbital variation analysis is a comparable survey duration (especially to alleviate the bias against longer-period outer main-belt objects). The use of only two visible-band filters\footnote{In fact mostly just one: 87\% of the $\sim$2 million small-body observations in this work are in Mould-$R$-band, 13\% in $g'$-band.} gives PTF an advantage over other ongoing surveys\footnote{To illustrate this point by comparison, of the $\sim$3 million small-body observations reported to the Minor Planet Center by \emph{Pan-STARRS 1} (PS1) as of mid-2012, $\sim$40\% are $w$-band (a wide-band filter covering most of the visible), while $g$-, $r$- and $i$-bands each represent $\sim$20\% of the PS1 data.}, since conversion between wavelength bands introduces uncertainty when object colors are unknown. Thus, multi-filter data makes absolute magnitude comparison between epochs (already complicated by uncertainties in spin amplitudes and phase functions) even more difficult. Finally, a photometric variation analysis would benefit from the inclusion of null-detections, which are not currently a product of our kd-tree harvesting method, but should be implementable with a reasonable amount of modification.

\section*{Acknowledgements}

This work is based on data obtained with the 1.2-m Samuel Oschin Telescope at Palomar Observatory as part of the Palomar Transient Factory project, a scientific collaboration between the California Institute of Technology, Columbia University, Las Cumbres Observatory, Lawrence Berkeley National Laboratory, the National Energy Research Scientific Computing Center, the University of Oxford, and the Weizmann Institute of Science (WIS). E.O. Ofek is incumbent of the Arye Dissentshik career development chair and is grateful for support via a grant from the Israeli Ministry of Science. O. Aharonson and E.O. Ofek wish to thank the Helen Kimmel Center for Planetary Science at WIS. S.R. Kulkarni and his group are partially supported by NSF grant AST-0507734. D. Polishook is grateful to the AXA research fund.

\label{lastpage}

\end{document}